\RequirePackage{fix-cm}

\documentclass[twocolumn]{svjour3} 
\smartqed 
\usepackage{latexsym}
\usepackage{times}
\usepackage{mathptmx} 

\usepackage{amsmath}
\usepackage{amssymb}
\usepackage{verbatim} 

\usepackage{accents}

\usepackage{subcaption}
\usepackage{graphicx}

\usepackage{enumerate}
\usepackage{hyperref}
\usepackage[utf8]{inputenc}
\usepackage{amsfonts}
\usepackage{amsbsy}
\usepackage{calrsfs}
\usepackage{mathtools}
\usepackage[dvipsnames]{xcolor}

\usepackage{etoolbox}
\patchcmd{\quotation}{\rightmargin}{\leftmargin 2.5em \rightmargin}{}{}

\DeclareMathOperator*{\minimize}{\ \ \ \text{\upshape minimize}\ \ \ }
\DeclareMathOperator*{\maximize}{\ \ \ \text{\upshape maximize}\ \ \ }
\DeclareMathOperator*{\subjectto}{\ \ \ \text{\upshape subject to}\ \ \ }

\usepackage{changepage}

\newcommand{\ubar}[1]{\underaccent{\bar}{#1}}

\usepackage{float}

\newcommand{\mymycolor}{blue}
\newcommand{\add}[1]{{#1}}
\newcommand{\addd}[1]{{#1}}
\newcommand{\adddd}[1]{\textcolor{\mymycolor}{#1}}

\usepackage{color}
\definecolor{forestgreen}{rgb}{0.33,0.61,0.34}
\definecolor{deepmagenta}{rgb}{0.8, 0.0, 0.8}
\definecolor{harvardcrimson}{rgb}{0.79, 0.0, 0.09}

\definecolor{amethyst}{rgb}{0.6, 0.4, 0.8}

\begin{document}

\title{Optimal Resource Allocation for Dynamic Product Development Process via Convex Optimization
}


\author{Chengyan Zhao \and
Masaki Ogura \and Masako Kishida
\and
Ali Yassine
}

\institute{Chengyan Zhao \at
Graduate School of Science and Technology, Nara Institute of Science and Technology, Ikoma, Nara 630-0192, Japan \\
\email{zhao.chengyan.za5@is.naist.jp.}
\and
Masaki Ogura \at
Graduate School of Information Science and Technology,
Osaka University, Suita, Osaka 565-0871, Japan \\
\email{m-ogura@ist.osaka-u.ac.jp.} 
\and
Masako Kishida \at
Principles of Informatics Research Division, National Institute of Informatics, Tokyo 101-8430, Japan\\
\email{kishida@nii.ac.jp.} 
\and 
Ali Yassine \at
Department of Industrial Engineering and Management, American University of Beirut, Beirut 1107-2020, Lebanon \\
\email{ali.yassine@aub.edu.lb.} 
}

\date{Received: date / Accepted: date}
\maketitle

\begin{abstract}
Resource allocation is an essential aspect of successful Product Development (PD). In this paper, we formulate the dynamic resource allocation \add{problem} of the PD process as a convex optimization problem. Specially, we build and solve two variants of this \add{problem}: the budget-constrained problem and the performance-constrained problem. \add{We use} convex optimization \add{as} a framework to optimally solve large problem instances at a relatively small computational cost. The solutions to both problems exhibit similar trends regarding resource allocation decisions and performance evolution. Furthermore, we show that the product architecture affects resource allocation, which in turn affects the performance of the PD process. By introducing centrality metrics for measuring the location of the modules and design rules within the product architecture, we find that resource allocation decisions correlate to their metrics. These results provide simple, but powerful, managerial guidelines for efficiently designing and managing the PD process. Finally, for validating the model and its results, we introduce and solve two design case studies for a mechanical manipulator and for an automotive appearance design process.

\end{abstract}

\section{Introduction}\label{intro}

Successful Product Development (PD) requires careful allocation of development resources. Allocating resources to various subsystems and modules within the product system requires a deep understanding of many complex interactions. These interactions arise from various sources; namely, due to the physical interdependencies between the different subsystems in the product itself (i.e., the product architecture), the arrangements of organizations that will carry out the development process (i.e., the social network behind the organization), and the structure of the development process (i.e., predecessor relationships between development activities)~\cite{Yassine2019a}. In particular, this paper is focused on obtaining an understanding of the product architecture and its role in resource allocation decisions during PD\add{\footnote{\add{Product architecture is not the only driver for resource allocation decisions. Other drivers, such as existing product lineup, competitive products, product demand and price, technological advancements, consumer taste changes, balancing the development portfolio, etc., may play an influential role \cite{Terwiesch2008}. However, we focus on product architecture since it is the central issue in our proposed model.}}}.

Product architecture is the scheme by which the functional elements of the product are arranged into physical chunks and by which the chunks interact~\cite{Ulrich1995}. Product architecture plays a significant role in every aspect of the product lifecycle from influencing how the product is designed, manufactured, marketed, experienced, serviced, and retired~\cite{Ulrich2008,Yassine2007}. Additionally, the product architecture has profound implications for many product behavior properties from robustness~\cite{Braha2007} to evolvability~\cite{Luo2015}. It also influences how resources are allocated in the PD process~\cite{Ogura2019,Yassine2016}. 

Product architecture is usually described by a continuum between an integral product architecture to a modular one. In integral architectures, the product functions are shared by product modules (i.e., physical elements), and in modular architectures, each function is delivered by a separate element or module. Thus, integrality creates interdependence between product elements or modules. This interdependency, in turn, results in complexity. That is, some of the interdependencies may not be known in advance, or their influence on product and PD process performance may also be unknown. Within this complex PD environment, several studies have argued that the product architecture may evolve from integral to modular~\cite{Christensen2002,Yassine2016}.

In this paper, we investigate how the product architecture may influence the resource allocation decision to various modules using an optimization framework. Using this framework, we investigate the tendency for product architectures to evolve form integral to modular architectures. The main objective of this paper is to check whether the location of a module within the product architecture can offer PD managers insights into optimal resource allocation decisions. 

Several authors have formulated and analyzed the PD problem by analogy to dynamic linear systems (e.g.,~\cite{Ogura2019,Smith1997,Yassine2003}). In their analysis, they assumed that all tasks in the design structure matrix (DSM) proceed in parallel, where the DSM is a matrix representation of the development network. At any iteration stage, one unit of work on one task results in a fraction of rework for the other dependent tasks during the next iteration stage. The dependency between tasks is captured by the numerical values in the DSM. As such, the work completed in a current design iteration is a linear function of the work completed in the previous design iteration, with the linear weights being the numerical values in the DSM. 

Other authors have used complexity theory to describe and analyze the PD process~\cite{Ahmad2018}. For instance, Braha and Bar-Yam~\cite{Braha2007} introduced the NK-based model and analysis of product development project networks. They showed how the underlying network topologies and statistical structural properties provide direct information about the functionality, dynamics, robustness, and fragility of these PD projects. Also, the authors in~\cite{Frenken2012} argued that modules could be optimized independently if interface standards between modules are left unchanged. Similarly, Luo~\cite{Luo2015} used the NK framework to show how different product architectural patterns can influence product evolvability.

Network analysis has also been used for analyzing PD project network~\cite{Batallas2006,Collins2009}\label{page:intro:chroho}. For example, the analysis of the network structure (i.e., statistical properties) for various software and hardware development projects in~\cite{Braha2007} revealed that these networks have both small world and scale free network patterns. Additionally, they demonstrated that complex design networks are highly robust to the failure of randomly selected design components, but weak for failures targeting specific components (such as hub components). Similarly, Sosa et al.~\cite{Sosa2011} found that the analysis of the network structure of complex product designs (particularly, the existence of hubs in the design network) impacts the quality of the product being developed.

More recently, the authors in~\cite{Yassine2016} have formulated the PD resource allocation problem as a nonlinear optimization problem. Furthermore, the authors proposed a dynamic model in which there are several investment runs (or rounds) during the PD process. This formulation allowed the investigation of several interesting hypotheses, including the impact of architecture on performance evolution from integral to modular systems.

In this paper, we offer a more efficient optimization approach based on convex optimization techniques, which would allow us to find the globally optimal allocation of development resources\footnote{\add{See \cite[Section~4]{Reich2004} for a related discussion on convex optimization in the context of engineering design.}}\label{citeReich2004}. In this direction, we first adopt a discrete-time linear system to represent the work transformation feature in the PD process. Then, we propose an optimization framework where the resource allocation problem of the PD process can be transformed into a convex optimization problem. \add{We then apply our framework to symmetric and synthetic product architectures to reveal the trends of the evolution of optimal investment.} Finally, \add{by analyzing real case studies with asymmetric PD architectures, we gain insights into} the resource allocation problem and provide a guide for designing and managing the PD process. 

The following sections are organized as follows. In Section~\ref{Pb_formulation}, we describe the work transition feature of the PD process by a discrete-time linear system. Then, we formulate the budget-constrained problem and the performance-constrained problem for optimal resource allocation. In Section~\ref{section_main_result}, we propose the framework that both the optimization problems can be transformed into the convex optimization problems. In Section~\ref{simulation}, we perform \add{the experiments and analysis} for \add{the optimal solution} of the PD process. Finally, we \add{carry out} two case studies to illustrate the result of this paper in Section~\ref{case_study}.

\section{Proposed model} \label{Pb_formulation}

In this section, we first review the dynamic model of the PD process proposed in~\cite{Yassine2016}. Then, from the perspective of system and control, we show that the work transition feature in the PD process can be expressed by a discrete-time linear system. Finally, we formulate the optimal resource allocation problem of the PD process as the budget-constrained optimization and the performance-constrained problem separately.

\subsection{Work transformation matrix}\label{WTM_process}

In PD, the product architecture is built not only by the constituent parts that define the product system (i.e., modules or components), but also by the interaction relationships between these parts (i.e., dependency structure)~\cite{Ulrich1995}. In this paper, we assume that the product architecture has been determined in the early design stage. That is, the modules and their dependency structure (i.e., design rules) have been established. In this situation, we focus on improving the performance of PD system through allocating the development resources to the various modules and design rules over \add{all} investment rounds (i.e., design iterations).

We start the problem formulation by reviewing the dynamic PD model presented in~\cite{Yassine2016}. Suppose that there are $n$ modules and $T$ investment rounds during the PD process, we let $P_i(k)$ represent the amount of the remaining work in the $i$th module after finishing the $k$th investment round. The remaining work of all modules is defined by the vector
\begin{equation*}\label{eq:workvectorvar}
P(k)=
\begin{bmatrix}
P_1(k) \\
\vdots \\
P_n(k)
\end{bmatrix}.
\end{equation*}
The \add{progress} of the PD process is evaluated by the sum of the remaining work in each module~\cite{Joglekar2001}, which implies that the less total remaining work, the higher performance the product system has. Thus, the \addd{inverse of the} sum of the remaining work from all modules can be adopted as a measurement of the performance of the PD system at each round, which is expressed by the following expression:
\begin{equation}\label{eq:performance}
\addd{1/}\sum_{i=1}^{n} P_i(k). 
\end{equation}
At each iteration stage, the module finishes a certain amount of remaining work, and sends/receives the produced work (i.e., a fraction of rework) to/from its dependent modules. To describe this work transformation process, we use a discrete-time linear system expressed by the following equation:
\begin{equation}\label{eq:worktrans}
\addd{P(k)=A_k(\phi_k, \gamma_k)P(k-1),~k=1, \dotsc, T,}
\end{equation}
where $A_k(\phi_k, \gamma_k)$ is the work transformation matrix (WTM), 
\begin{equation*}
\phi_k=\{\phi_{1,k}, \dotsc, \phi_{n,k}\}	
\end{equation*}
\add{is the set of} the work completion rate of modules, and 
\begin{equation*}
\gamma_k=\{\gamma_{ij,k}\} \quad (i,j=1, \dotsc, n,\ i \neq j)
\end{equation*}
\add{is the set of} the updated value of design rules. \add{The value of the inter-module variable~$\gamma_{ij,k}$ represents the work flow strength from module $i$ to $j$ at $k$th investment round, i.e., at round~$k+1$, the accumulated produced work to module $j$ is the sum of the multiplication of the remaining work $P_i(k)$ on module $i$ and $\gamma_{ij, k}$.}\label{gamma}
For an established product architecture, the performance of the product system can be further improved by investing in both modules (i.e., determining the work completion rate in \add{each} iteration stage) and design rules (i.e., reducing the dependency strength between two modules). We assume that $\phi_k, \gamma_k$ can be tuned within the following intervals:
\begin{equation*}
0<\ubar{\phi}_{i, k} \leq \phi_{i, k} \leq \bar{\phi}_{i, k},\quad 0<\ubar{\gamma}_{ij, k} \leq \gamma_{ij, k} \leq \bar{\gamma}_{ij, k},
\end{equation*}
where $\bar{\phi}_{i, k}, \bar{\gamma}_{ij, k}$ are the initialized parameter values in \eqref{eq:worktrans}, and $\ubar{\phi}_{i, k}, \ubar{\gamma}_{ij, k}$ are \add{the upper and lower bounds of the parameters, respectively}. For the PD process with multiple investment rounds, the authors in~\cite{Yassine2016} showed that unlike the memoryless feature in the investment for the modules, the investment in the design rules \add{for reducing the work flow strength} has a cumulative effect. That is, $\gamma_{ij,k}$ at the $k$th round is updated to include the values of the design rules that resulted from the investment \add{prior to} the $k$th round. Therefore, the updated \add{values} of the design rules at the $k$th iteration $\gamma_{ij,k}$ is the multiplication of the updated value of the design rules from the $1$st to the $k$th \add{round}, which is expressed as $\prod_{\ell=1}^k \gamma_{ij, \ell}$. The specific form of the matrix~$A_k(\phi_k, \gamma_k)$ is given by
\begin{equation*}
A_k(\phi_k, \gamma_k)=
\begin{bmatrix}
\phi_{1, k} &\prod_{\ell=1}^k \gamma_{12, \ell} &\cdots &\prod_{\ell=1}^k \gamma_{1n, \ell} \\ 
\prod_{\ell=1}^k \gamma_{21, \ell} &\phi_{2, k} &\cdots &\prod_{\ell=1}^k \gamma_{2n, \ell} \\ 
\vdots &\vdots & \ddots &\vdots\\
\prod_{\ell=1}^k \gamma_{n1, \ell} &\prod_{\ell=1}^k \gamma_{n2, \ell} &\cdots &\phi_{n, k}
\end{bmatrix}
.
\end{equation*}

Suppose that we can use the development resources to update the value of $\phi_k, \gamma_k$. That is, we can use the resources to tune the work completion rate $\phi_k$ and the dependency strength $\gamma_k$. Moreover, we assume that there is an associated cost $f_i(\phi_{i, k})$ for tuning the value from $\bar{\phi}_{i, k}$ to $\phi_{i, k}$. Likewise, $g_{ij}(\gamma_{ij, k})$ is the cost for tuning the value $\bar{\gamma}_{ij, k}$ to $\gamma_{ij, k}$. Then, the total cost at the $k$th investment round equals
\begin{equation}\label{cost}
B_k(\phi_{k}, \gamma_{k})=\sum_{i=1}^{n} f_i(\phi_{i, k})+\sum_{i=1}^{n}\sum_{i\neq j}g_{ij}(\gamma_{ij, k}).
\end{equation}


Form the perspective of \add{the} project manager, \add{it is imperative to optimally allocate} the development resources to obtain the \add{maximum benefit}, especially when a huge project is carried out. However, making the optimal resource allocation strategy for thousands of decision variables just by experience and intuition seems not very effective. Thus, a mathematical programming formulation for finding the optimal investment strategy is essential.

\subsection{Optimization problem}\label{problem_formulation}

As mentioned in Section~\ref{WTM_process}, at each iteration, PD managers can use a certain amount of development resources to improve the performance of the product system. Particularly, the resources can be allocated on a module for tuning its work completion rate or on the design rule for reducing the dependency strength between two modules. Assuming that given a set of budgets for each investment round during the whole development process, how should we make the allocation strategy to minimize the total remaining work of the PD process? Based on this question, we formulate the budget-constrained problem as follows: 
\begin{problem}[Budget-constrained optimization]\label{pb1}
\textup{Assume \\ that, given $P(0)$, there are $T$ investment rounds with the corresponding budgets $\bar{B}_k >0$ ($k=1, \dotsc , T$) for resource allocation during the PD process, as well as the cost functions $f_i(\phi_{i, k})$ and $g_{ij}(\gamma_{ij, k})$. Find a sequence of decision variables for allocating the resources in modules $\phi=\{\phi_{i, k}\}_{k=1}^T$~$(i=1, \dotsc , n)$ and design rules $\gamma=\{\gamma_{ij, k}\}_{k=1}^T (i, j=1, \dotsc , n, \ \ i \neq j)$ to
\addd{maximize the performance measure~\eqref{eq:performance}.}} 
\end{problem}

\textup{Mathematically, we formulate the budget-constrained \\ problem as:}
\begin{subequations}\label{pb1:}
\begin{align}
\addd{\maximize_{\phi, \gamma}} & \addd{1/}\sum_{i=1}^{n}P_i(T) \label{pb1:A}
\\
\subjectto & B_k(\phi_{k}, \gamma_{k}) \leq \bar{B}_k, \label{pb1:B}
\\
&\add{0<\ubar{\phi}_{i, k} \leq \phi_{i, k} \leq \bar{\phi}_{i, k},}\label{pb1:D}\\ &\add{0<\ubar{\gamma}_{ij, k} \leq \gamma_{ij, k} \leq \bar{\gamma}_{ij, k},}~\add{k=1, \dotsc ,T.}\label{pb1:C} 
\end{align}
\end{subequations}
For the budget-constrained problem, our goal is to make the optimal resource allocation strategy to maximize the PD \add{performance}. However, PD managers also face another \add{challenge} when the manager
plans to meet prescribed target on the remaining work at $T$. \add{The question becomes:} how to make the resource allocation \add{decision} to minimize the \add{cost of meeting the prescribed performance requirement}? In this case, \add{we formulate} the performance-constrained problem \add{as follows}:

\begin{problem}[Performance-constrained optimization]\label{pb2}\\
\textup{Assume that, given $P(0)$, there are $T$ investment rounds and the prescribed 
\addd{performance requirement $\ubar{\pi} > 0$,} 
as well as the cost functions $f_i(\phi_{i, k})$ and $g_{ij}(\gamma_{ij, k})$. Find a sequence of decision variables for allocating the development resources in modules $\phi=\{\phi_{i, k}\}_{k=1}^T$~$(i=1, \dotsc , n)$ and design rules $\gamma=\{\gamma_{ij, k}\}_{k=1}^T~(i,j=1, \dots, n,~i \neq j)$ to minimize the total investment while satisfying the \addd{requirement on the performance~\eqref{eq:performance}}.}
\end{problem}

\textup{As in \eqref{pb1:}, we can mathematically build the performance-constrained problem as the following:}
\begin{subequations}\label{pb2:}
\begin{align}
\minimize_{\phi, \gamma} & \sum_{k=1}^{T}B_k(\phi_{k}, \gamma_{k}) \label{pb2:A}
\\
\subjectto & \addd{1/}\sum_{i=1}^{n}P_i(T) \addd{\geq \ubar{\pi}},\label{pb2:B}
\\
&\add{0<\ubar{\phi}_{i, k} \leq \phi_{i, k} \leq \bar{\phi}_{i, k},}\\ &\add{0<\ubar{\gamma}_{ij, k} \leq \gamma_{ij, k} \leq \bar{\gamma}_{ij, k}.}~\add{k=1, \dotsc ,T.} \label{pb2:C}
\end{align}
\end{subequations}

\addd{Notice that the performance constraint~\eqref{pb2:B} is equivalent to the following constraint on the sum of the remaining work: 
\begin{equation*}
\sum_{i=1}^{n}P_i(T) \leq \bar{P},
\end{equation*}
where
\begin{equation*}
\bar{P} = 1/\ubar{\pi}
\end{equation*}
is regarded as the maximum allowable amount of the remaining work left at the end of the investment.}

The difficulty of solving \add{the budget-constrained problem and the performance-constrained problem} mainly stems from the nonlinearity of the functions \eqref{pb1:A}, \eqref{pb2:A} and constraints \eqref{pb1:B}, \eqref{pb2:B}.
That is, the budget-constrained problem and \add{the performance-constrained} problem become nonlinear optimization problems. Although there are some numerical solutions for this case based on heuristic methods~\cite{Alcaraz2003a,Boctor1996}, such techniques can cause the solution to be trapped in a local optimal point. Moreover, the computation cost of the heuristic solver grows rapidly with the increase in problem size (i.e., the number of modules, design rules and the investment rounds). Thus, there exists a need for developing a computation framework that can deliver the optimal solution for \add{a relatively large size of the resource allocation problem.}

\begin{figure*}[tb]
\centering
\includegraphics[height=4cm]{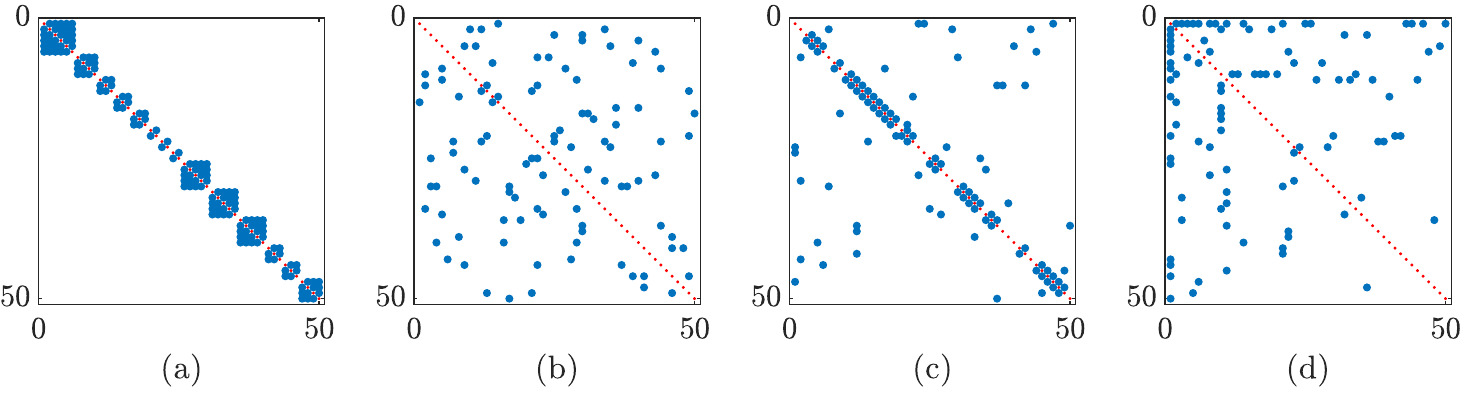} 
\caption{Four DSM architectures. All with $50$ modules and $100$ design rules: (a) block-diagonal network, (b) Erd\H{o}s-R\'{e}nyi network (Random), (c) Watz-Strogatz network (Small world), (d) Barab\'{a}si-Albert network (Scale free). The diagonals in the DSM represent the location of modules, and the off-diagonals show the dependencies between modules.} 
\label{figDSM} 
\end{figure*}

\section{Solution using convex optimization}\label{section_main_result}

In this section, we present an optimization framework for efficiently solving the budget-constrained problem and the performance-constrained problem. Under the relatively mild assumption on the cost function, we \add{can} show that problems can be transformed into convex optimization problems. Let us begin with reviewing the definition of posynomials with the following:
\begin{definition}[\cite{Boyd2004}]\label{def1}
Let $v=\{v_1$, $\dotsc$, $v_n$\} denote $n$ real positive variables.
\begin{enumerate}
\item We say that a real function $g(v)$ is a {\it monomial} if there exist $c>0$ and a set of real numbers $a_1, \dotsc, a_n$ such that $g(v) = c v_{\mathstrut 1}^{a_{1}} \dotsm v_{\mathstrut n}^{a_n}$.
\item We say that a real function $f(v)$ is a {\it posynomial} if $f$ is a sum of monomials of $v$.
\end{enumerate}
\end{definition}\label{explicyt}

To precisely model the cost features, the nonlinearity from the practical problem can not be ignored. From Definition~\ref{def1}, we can see that the posynomials are nonlinear functions which can be used for fitting real data from practical PD problems. For specific techniques to fit posynomials to real data, we refer readers to~\cite{Boyd2007a}. Also, the following lemma shows the convexity property of posynomials, which is essential in transforming the budget- and performance-constrained problems into convex optimization problems.

\begin{lemma}[\cite{Boyd2007a}]\label{lem1}
{\textup{If $f$ is a posynomial, then, the function} $x \mapsto \log f(\textup{exp}[x])$
\textup{is convex.}}
\end{lemma}

As mentioned earlier, the nonlinearity of the real data can be fitted by posynomials. From Definition~\ref{def1}, we can see that the range of the posynomials is in the nonnegative number field. However, in practice, real data may run out of the nonnegative area. Thus, normalizing the range of the cost function to the nonnegative field is necessary (i.e., adjust the minimum value of the cost function larger than $0$). \add{For} ensuring the non-negativity of the cost function, we assume a specific structure~\add{\cite{Ogura2020}} on the cost function:
\begin{equation*}\label{cost_assumption}
\begin{aligned}
f_i(\phi_{i, k}) &= f_i^+(\phi_{i, k})-f_i^+(\bar{\phi}_{i, k}),
\\
g_{ij}(\gamma_{ij, k}) &= g_{ij}^+(\gamma_{ij, k})-g_{ij}^+(\bar{\gamma}_{ij, k}).
\end{aligned}
\end{equation*}
The essential part of the cost function is the first term $f_i^+(\phi_{i, k})$, while the second term $(-f_i^+(\bar{\phi}_{i, k}))$ is for normalizing the cost function as $f_i(\bar{\phi}_{i, k})=0$, similarly for $g_{ij}(\gamma_{ij, k})$, which means that the zero investment yields no cost. 

\add{The resulting optimization problems \eqref{pb1:} and \eqref{pb2:} are not trivial to solve directly because the nonlinearity in the work transformation process function \eqref{pb1:A}, \eqref{pb2:B} and the resource cost functions \eqref{pb1:B}, \eqref{pb2:A}. Although there exist heuristic optimization methods that can solve this problem, the solution is local optimal due to the constraint of the algorithm. In a complex product development project that contains hundreds of modules and design rules, finding the global optimal resource allocation strategy can bring great benefit for the company and shareholders. Therefore, it is necessary to establish an efficient computation framework for obtaining the global optimal solution to the problems \eqref{pb1:} and \eqref{pb2:}.}

\add{The following theorem allows us to overcome the difficulty and solve the budget-constrained problem and the performance-constrained problem via convex optimization and is the main theoretical result of this paper. The proof of the theorem is presented in the Appendix.}

\begin{theorem}
Problems~\ref{pb1} and~\ref{pb2} reduce to convex optimization problems.
\end{theorem}

\begin{figure}[tb]
\centering
\includegraphics[height=5.2cm]{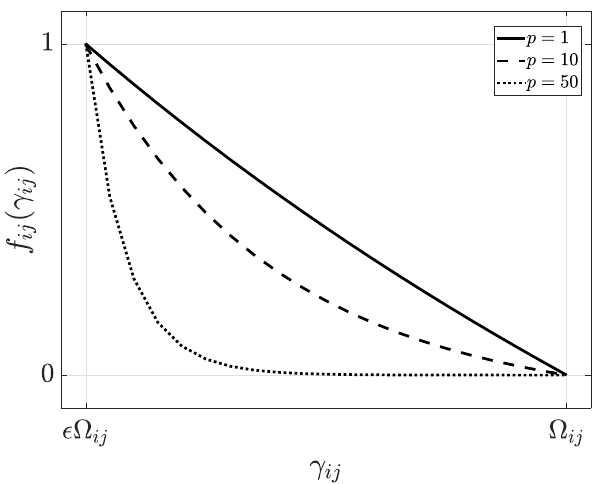}
\caption{Three cost functions with $p$ = 1, 10, and 50. $f_{ij}(\Omega_{ij})=0$ represents that no resource is allocated, where $\Omega_{ij}$ denotes the initial value of the certain entry in WTM. $f_{ij}(\epsilon\Omega_{ij})=1$ indicates the upper bound of the allocated resources where we can to obtain the fully improved value.}
\label{figcostfun} 
\end{figure}

\section{Experimental setup, analysis and discussion of results}\label{simulation}

In this section, we show the effectiveness of the proposed framework by solving relatively large-size PD problems with different product architectures. Furthermore, by investigating the solution, we reveal the trends, structure, and relationship of the decision variables. In Section~\ref{sim_networks}, we introduce four typical DSM architectures embedded in our simulation experiments. In Section~\ref{sim_costfunction}, we give the specific form of the cost function. Then, in Section~\ref{sim_prob1}, we present the optimal solution of the budget-constrained problem, perform its analysis, and discuss the results. Likewise, in Section~\ref{sim_prob2}, the optimal solution of the performance-constrained problem is demonstrated and discussed. In Section~\ref{sim_result_compare}, we \add{statistically} investigate the impact of product architecture on \add{optimal resource allocation.} 

\subsection{DSM architecture}\label{sim_networks}

As mentioned earlier, the design structure matrix (DSM) is a matrix representation of the development network having a particular architecture~\cite{Jan2007}. For this reason, the DSM architecture in our experiment is determined by the following network models: the block-diagonal~\cite{Jan2007}, the Erd\H{o}s-R\'{e}nyi (random)~\cite{erdos59a}, the Watz-Strogatz (small world)~\cite{Watts1998a} and the Barab\'{a}si-Albert (scale free)~\cite{Barabasi1999} graphs. 

Fig.~\ref{figDSM} shows the four DSM architectures used in this paper. On one end, the block-diagonal network represents a typical modular architecture, where the dependencies between modules are divided into dependent groups (no interactions between the groups), and the modules in each group are fully dependent (see Fig.~\ref{figDSM} (a)~\cite{Yu2007}). Alternatively, the Watz-Strogatz network and the Barab\'{a}si-Albert network represent the other extreme, called integral architecture. The Watz-Strogatz network in Fig.~\ref{figDSM} (c) shows the small world property, where most modules dependencies are local, but few dependencies exist between the distant modules~\cite{Watts1998a,Yassine2016}. The Barab\'{a}si-Albert network in Fig.~\ref{figDSM} (d) illustrates the preferential attachment feature of the PD project. The project starts with few modules and as the design process unravels the new modules are linked to the old modules~\cite{Barabasi1999,Yassine2016}. We adopt the Erd\H{o}s-R\'{e}nyi network in Fig.~\ref{figDSM} (b) for randomly setting the dependency structure in the DSM, which serves as a benchmark to other patterned DSM architectures.
For other DSM architectures, we refer the readers to~\cite{Braha2004,Braha2004a,Yassine2016}. 



\subsection{Cost function}\label{sim_costfunction}

As mentioned in Section~\ref{WTM_process}, the resources allocated on the modules and design rules result in a reduction of \add{the parameters in~\eqref{eq:worktrans}}. Based on this, we claim that the cost function should be a decreasing function, and satisfy Definition~\ref{def1}. Thus, we use the following cost function:
\begin{equation}\label{cost:example}
f_{ij\add{, k}}(\gamma_{ij\add{, k}})=c_{ij}\left(\frac{1}{(\gamma_{ij\add{, k}})^p}-\frac{1}{(\Omega_{ij\add{, k}})^p}\right),
\end{equation} 
where \add{$\gamma_{ij, k}$}~($i,j=1, \dots, n,~i \neq j$) is the updated value of the parameter in the WTM, $p$ is a positive \add{real} number for tuning the shape of the concerned cost function, and $c_{ij}$, $\Omega_{ij}$ are positive numbers for fitting the value of the cost function to satisfy Definition~\ref{def1}. Then, we make the following assumption \add{to show} the diminishing return property, which ensures the convexity of the cost function as well. Suppose that there is a fixed increment $\epsilon_{ij}>0$ on $\gamma_{ij}$, and let $\Delta f_{ij}(\gamma_{ij})=f_{ij}(\gamma_{ij}-\epsilon_{ij})-f_{ij}(\gamma_{ij})$ represent the cost for tuning $\gamma_{ij}$ to $\gamma_{ij}-\epsilon_{ij}$. The diminishing return property means that the parameter tuning cost $\Delta f_{ij}(\gamma_{ij})$ increases with $\gamma_{ij}$, and also implies the convexity of $f_{ij}$. 
In practice, the parameters of the cost function are carefully assigned by the managers and the work teams (e.g., see~\cite{Stellman2005,Yassine2016}).
Fig.~\ref{figcostfun} shows three realizations of the cost function under different values of $p$.

\subsection{Analysis and discussion of the budget-constrained problem}\label{sim_prob1}

\begin{figure*}[tb]
\centering 
\includegraphics[width=.98\linewidth]{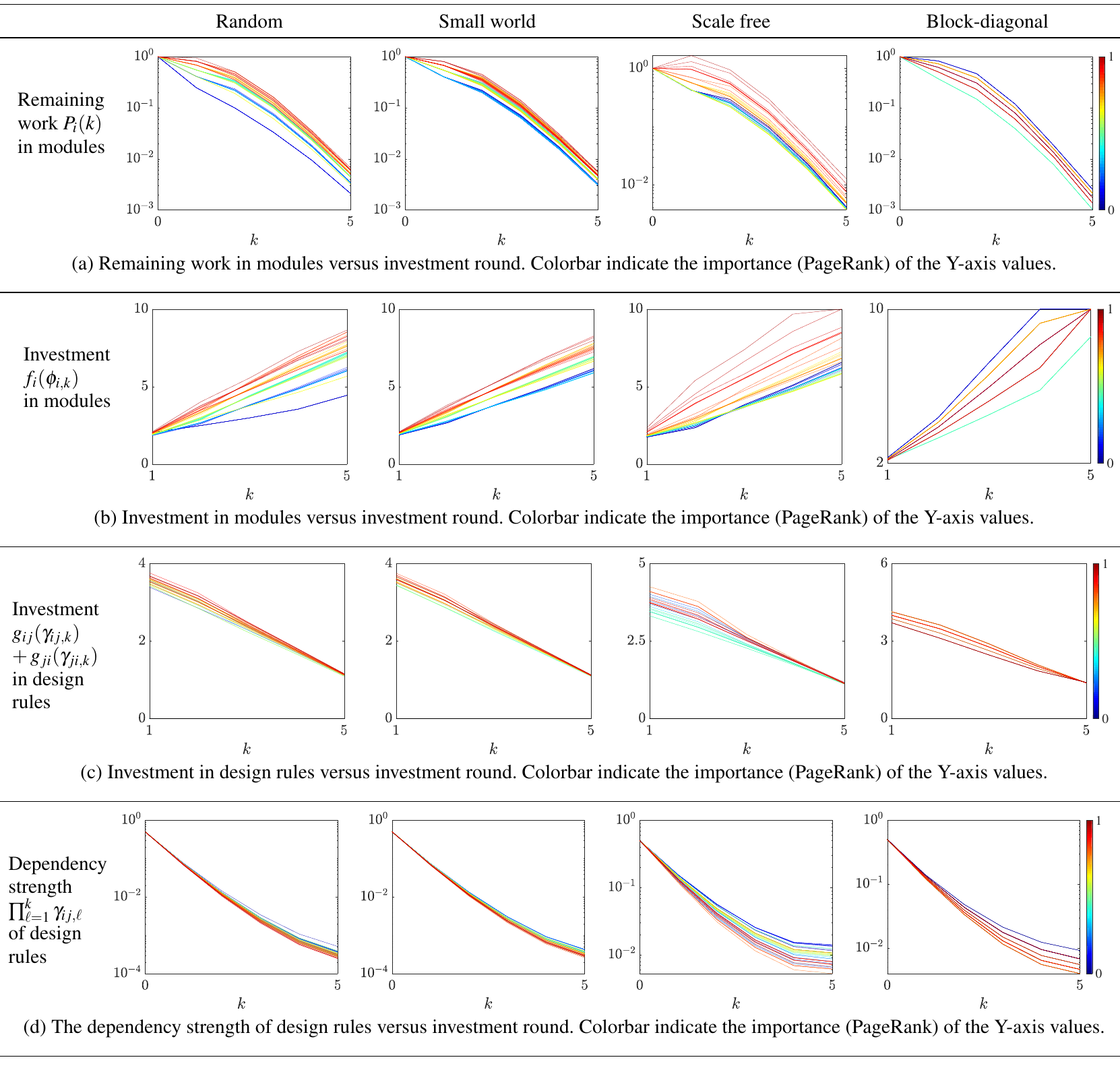}
\caption{The optimal solution of \add{the budget-constrained problem. Color lines distinguish the importance of modules/design rules via the PageRank (the importance is fixed from $0$ to $1$, see the colorbar on the right side). }} 
\label{result_budget_constraint} 
\end{figure*}

In this subsection, we optimally solve the budget-constrained problem through our proposed framework. Then, we investigate the evolution of decision variables during the budget-constrained PD process. Finally, we introduce the centrality metrics for measuring the importance of modules and design rules, and study whether the allocated resources or the remaining work in each module or design rule correlates with its centrality.

In this simulation experiment, for testing the effectiveness of solving a relatively large scale PD problem~\cite{Baldwin2000}, we produce the DSMs of size $50$ and hold the total number of dependencies to $100$ for each DSM architecture. We set the number of investment rounds $T=5$, and the budget $\bar{B}_k=300$ for each investment round. For initializing the parameters of the WTM, we unify $\bar{\phi}_{i, k}=0.5$~$(k=1,\dotsc, 5, i=1,\dotsc, 50)$ and $\bar{\gamma}_{ij, k}=0.05$~$(i,j=1,\dotsc, 50, i \neq j)$ for all the experiments. For all the cost functions, we unify the parameters with $c_{ij}=1$, $p=1$, $\Omega_{ij}=1$, and $\epsilon=0.1$, which indicates that the $\phi$ and $\gamma$ can be updated between $[0$$\%-90$$\%]$ of the initial value. From the parameter initialization, we can see that the values of $\phi_{i, k}$ and $ \gamma_{ij, k}$ can be tuned within the intervals $[0.05, 0.5]$ and $[0.005, 0.05]$, respectively. We \add{conduct} the experiments with the selected DSM architectures in Fig~\ref{figDSM}, and \add{observe} the following response variables: the remaining work in modules, the investment in the modules and the design rules, and \add{the dependency strength between modules.} 

For problem solving, we adopt the commonly used off-the-shelve software for convex optimization problem: \textsf{fmincon} routine in MATLAB. From the \add{experiment setup}, we can see that the total number of the decision variables is $(50+100)\times 5=750$, \add{which reaches the standard size of large PD process.} Through running the experiment on the desktop with common configuration (i.e., Intel Core i7-7700 and 8GB memory), the average time for solving the optimization problem is $210$ minutes, which illustrates that our framework \add{is capable for} solving \add{the much larger-scale} problems.

\begin{figure}[tb]
\centering
\includegraphics[width=.8\linewidth]{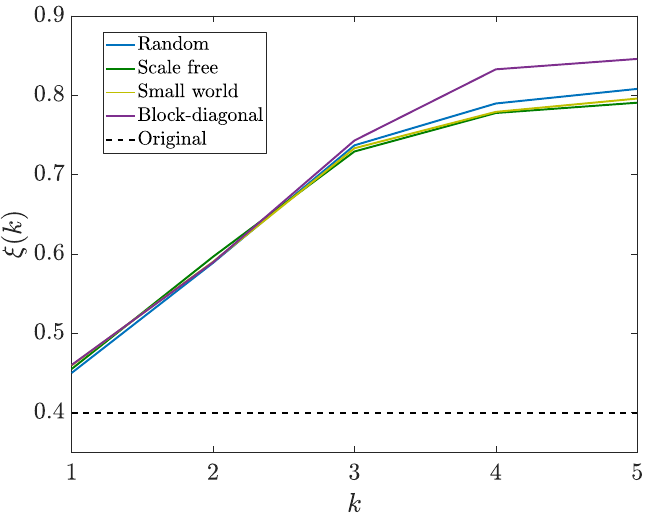}
\caption{\add{Evolution of the completion rate} \add{in} the budget-constrained problem.}
\label{fig:performance} 
\end{figure}

\begin{figure}[tb]
\centering
\includegraphics[width=.98\linewidth]{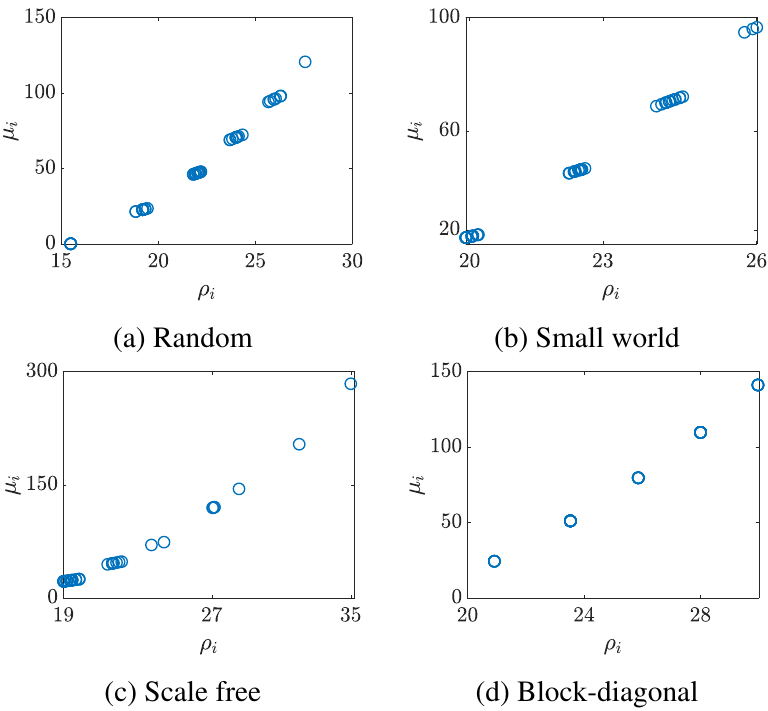}
\caption{{The correlation between the investment in module and the total investment in its dependent design rules in the optimal solution of the budget-constrained problem.}} 
\label{result_budget_constraint:spinout} 
\end{figure}

\begin{figure*}[tb]
\centering 
\includegraphics[width=.69\linewidth]{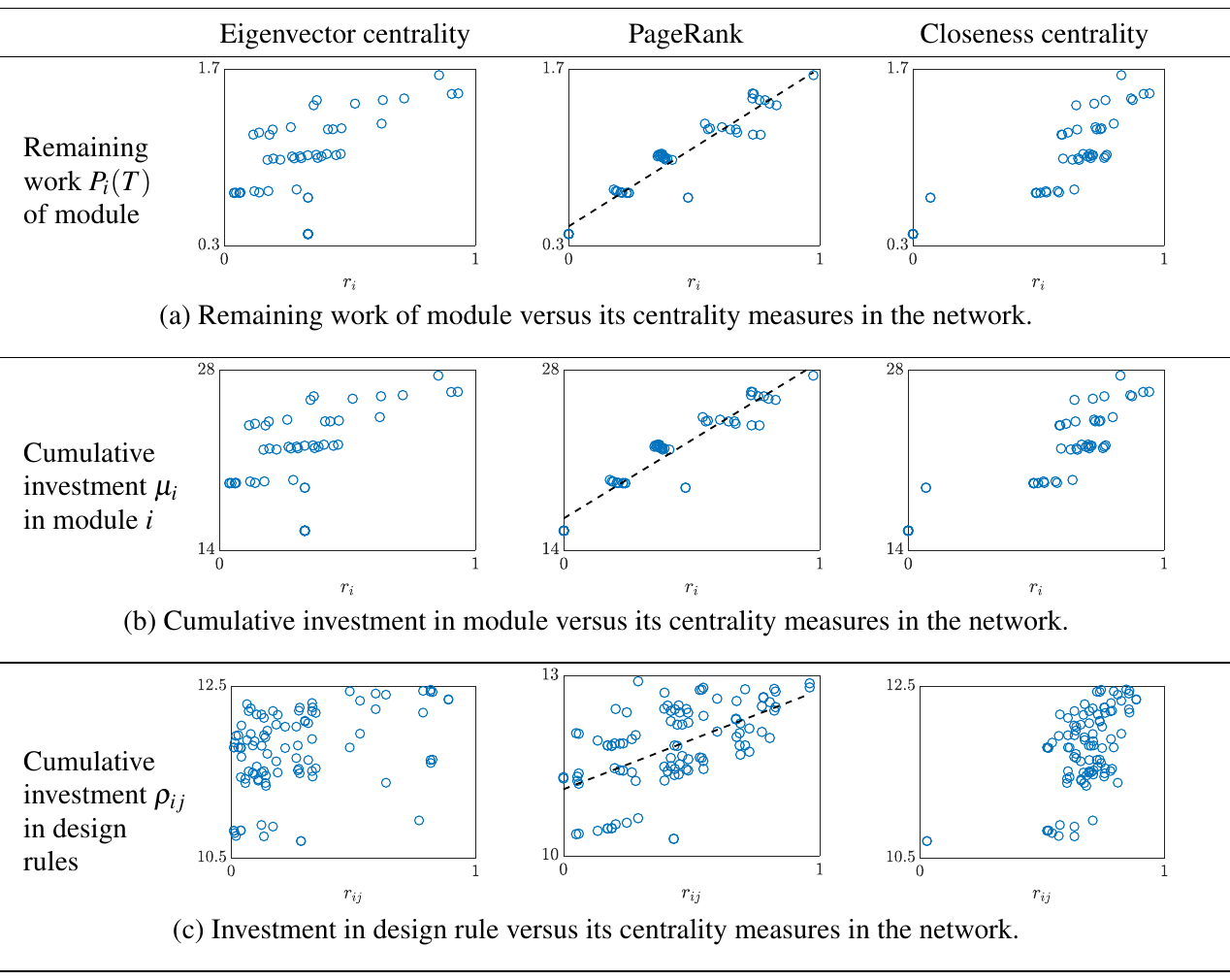}
\caption{The remaining work, investment in modules and design rules of Problem~\ref{pb1} versus their centrality measures in the Erd\H{o}s-R\'{e}nyi (random) network. Dash line: Linear regression line.} 
\label{fig_rand_10_budget_scatter} 
\vspace{5mm}
\includegraphics[width=.69\linewidth]{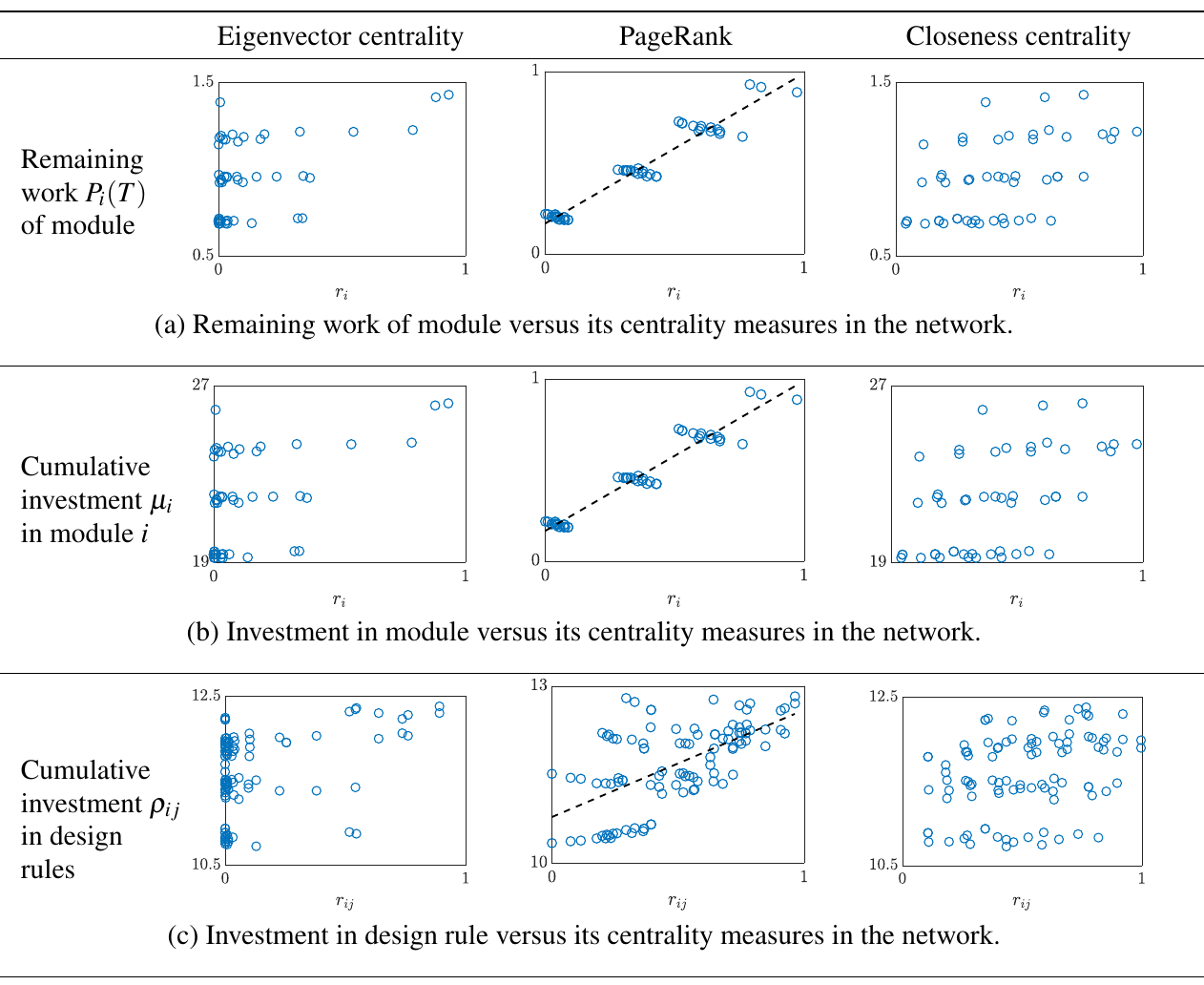}
\caption{The remaining work, investment in modules and design rules of Problem~\ref{pb1} versus their centrality measures in the Watz-Strogatz (small world) network. Dash line: Linear regression line.} 
\label{fig_smallworld_10_budget_scatter} 
\end{figure*}

\begin{figure*}[tb]
\centering 
\includegraphics[width=.69\linewidth]{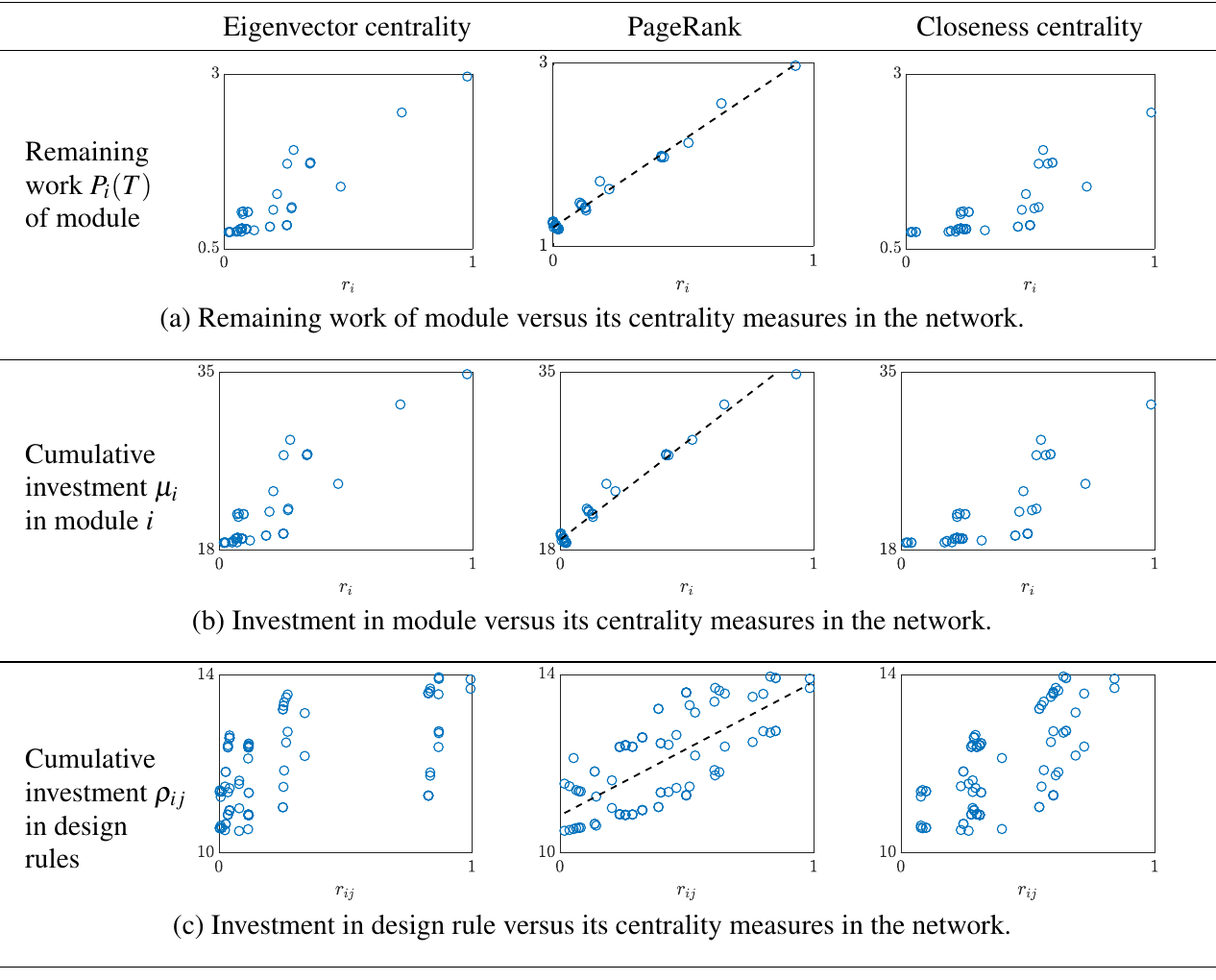}
\caption{The remaining work, investment in modules and design rules of Problem~\ref{pb1} versus theirr centrality measures in the Barab\'{a}si-Albert (scale-free) network. Dash line: Linear regression line. } 
\label{fig_scalefree_10_budget_scatter} 
\vspace{5mm}
\includegraphics[width=.69\linewidth]{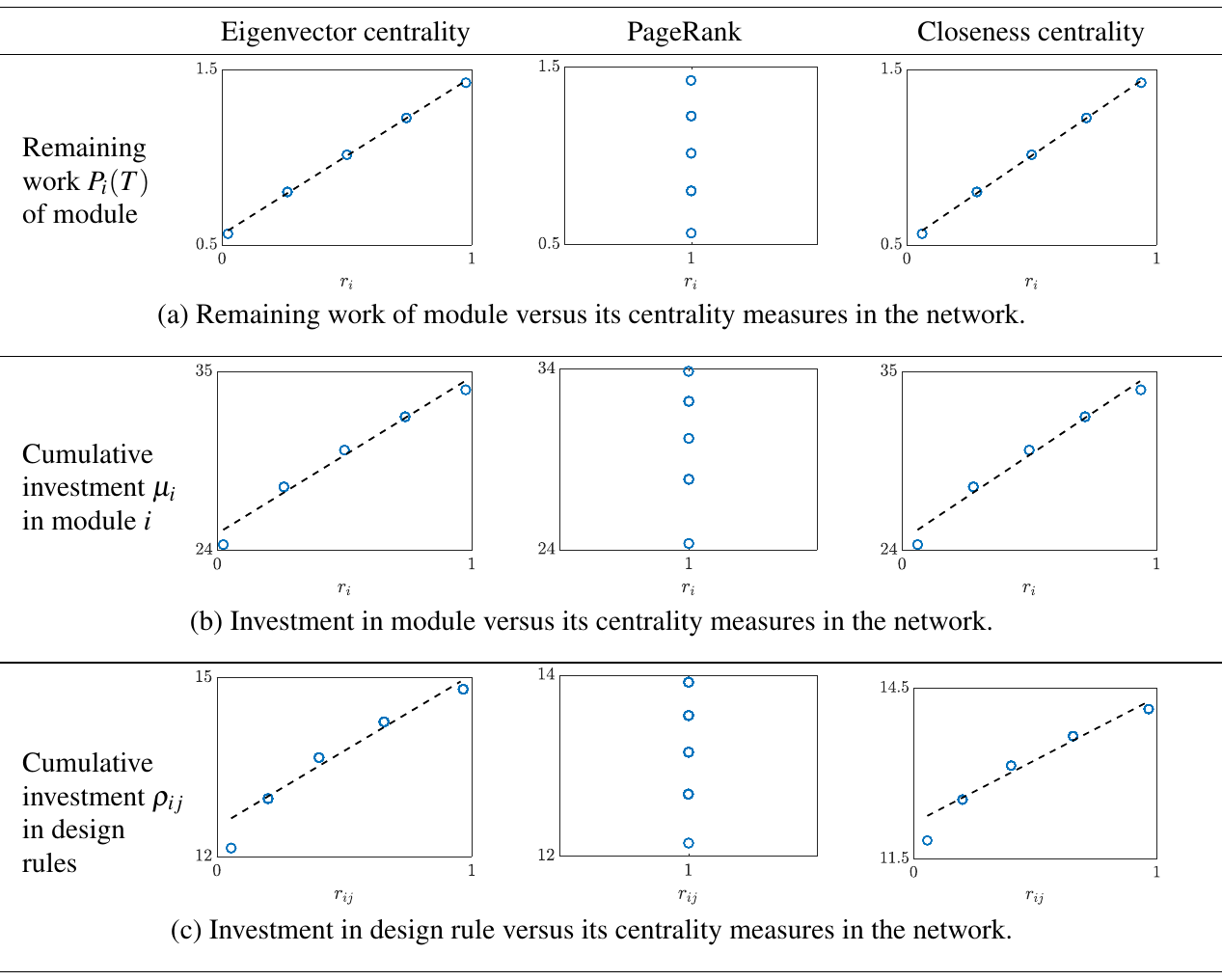}
\caption{The remaining work, investment in modules and design rules of Problem~\ref{pb1} versus their centrality measures in the Block-diagonal network. Dash line: Linear regression line.}
\label{fig_block_10_budget_scatter} 
\end{figure*}

Fig.~\ref{result_budget_constraint} shows the evolution of the resource allocation variables ($\phi_{i, k}$, $\gamma_{j i, k}$), the remaining work $P_i(k)$ and the dependency strength of design rules in solving the budget-constrained problem. \add{Particularly, Fig.~\ref{result_budget_constraint} (a) shows the decreasing trends of the remaining work in each module, which indicates that the PD process is in progress. However, this experiment contains more than $1$ investment round, so just by observing the remaining work from Fig.~\ref{result_budget_constraint} (a) is not easy to distinguish the effect of the investment at each round. Thus, we define the completion rate $\xi(k)$ for each investment round by}
\begin{equation*}
\xi(k)=\frac{\sum_{i=1}^{n}P_i(k)-\sum_{i=1}^{n}P_i(k+1)}{\sum_{i=1}^{n}P_i(k)}.
\end{equation*}
From Fig.~\ref{fig:performance}, we can observe the evolution of the \add{completion rate} during the process, where the performance of product system is \add{monotonically} improved with successive investment. Compared with the experiment with no investment (dashed line), we can derive that the PD process is accelerated with the allocated resources. As seen in Fig.~\ref{fig:performance}, with the investment in progress, the performance reaches its saturation. We also notice that the limits of performance are different \add{for} the DSM architectures, which implies that the DSM architecture must be taken into consideration for further studying the performance.
\add{We note that the high performance of the block-diagonal network is consistent with the finding in \cite{Braha2004} that real design networks have a nested hierarchical network structure.}\label{page:consistency}

Figs.~\ref{result_budget_constraint} (b) and~\ref{result_budget_constraint} (c) show the evolution of investment in modules and design rules, where we can see that the investment in modules increases, while the investment in design rules decreases. This phenomenon is in line with the result in~\cite{Yassine2016} that there is a shift in resource allocation from design rules to modules as the development process progresses. Moreover, it is worth noting that the modular architecture consumed more resources on the modules compared with design rules, while integral architecture consumed more resources on the design rules. Thus, the evolution of the decision variables also confirms that the product architecture evolves from integral to modular as the product matures. In Fig.~\ref{result_budget_constraint} (d), we can see that the dependency values of the design rules tends to $0$, which indicates that the dependency strength between modules is reduced or nearly eliminated by successive investment in design rules.

Next, we carry out a further investigation on whether there is a relationship between the \add{optimal} cumulative investment in a specific \add{module and its related design rules}. \add{We define the cumulative allocated resources on the module during the process by 
\begin{equation*}
\mu_i = \sum_{k=1}^{T}f_i (\phi_{i, k})
\end{equation*}
and its related design rules
\begin{equation*}
\rho_i = \sum_{k=1}^{T}\left(\sum_{j=1}^{n}g_{i j}(\gamma_{i j, k})+\sum_{j=1}^{n}g_{ji}(\gamma_{ji, k})\right).
\end{equation*}
We also define the cumulative allocated resources on design rules by}
\begin{equation*}
\add{\rho_{ij} = \sum_{k=1}^{T}\left(
g_{i j}(\gamma_{i j, k})
+
g_{ji}(\gamma_{ji, k})
\right).}
\end{equation*}
Fig.~\ref{result_budget_constraint:spinout} shows a positive correlation between the investment in the modules and its related design rules. This observation can be used as a managerial guideline for resource allocation: if a module is assigned with \add{a certain amount of resources}, then corresponding amount of resources must be allocated to its related design rules.

\begin{table*}[!tb]
\centering
\caption{Pearson correlation analysis for the budget-constrained problem (Figs.~\ref{fig_rand_10_budget_scatter}-\ref{fig_block_10_budget_scatter}) }\label{centrality_corr_spearman}
\includegraphics[width=.725\linewidth]{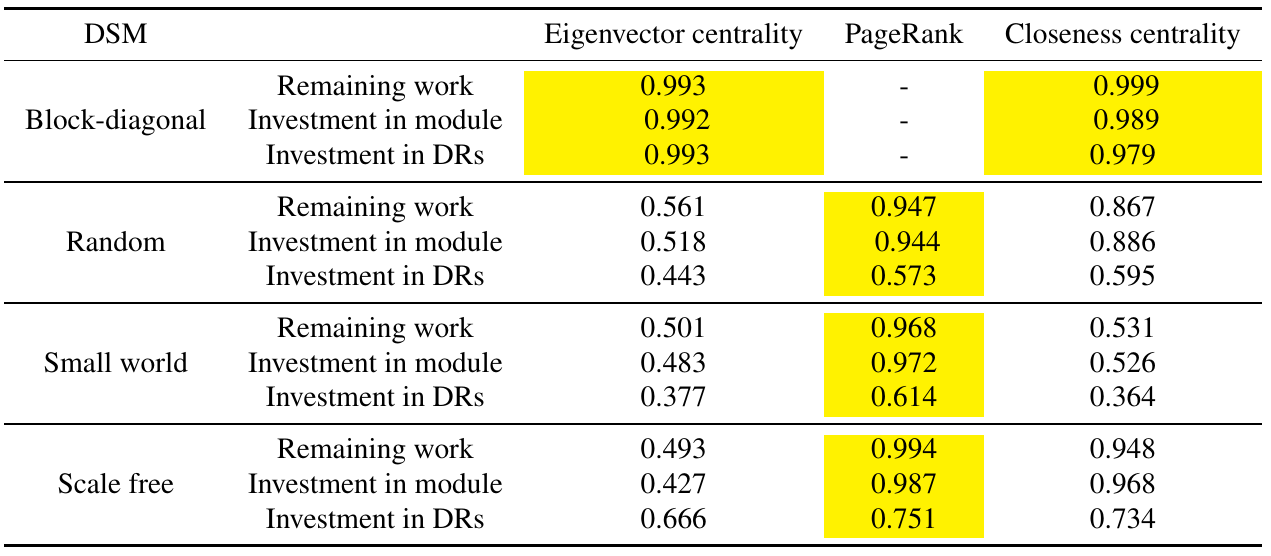}
\end{table*}

After discussing the results in Fig.~\ref{result_budget_constraint}, we introduce the centrality metrics (i.e., importance measures) for the modules and design rules to investigate whether there is a relationship between the investment in module/design rule and its centrality. For describing the importance of the modules and design rules, we adopt three centrality metrics~\cite{Newman2010}: the Eigenvector, the PageRank, and the Closeness centrality. \add{For simplicity, we normalize each centrality metric to $1$.} \add{Throughout the paper, we let the centrality metric of the $i$th module be denoted by $r_i$, and the centrality metric of a design rule between the $i$th and the $j$th module be denoted by $r_{i j}$, respectively. Then, Figs.~\ref{fig_rand_10_budget_scatter}-\ref{fig_block_10_budget_scatter} show the the dependence on the centrality measures of remaining work~$P_i(T)$, cumulative resource allocation $\mu_i$ on the modules, and cumulative resource allocation $\rho_{i j}$ on the design rules.} 
From Figs.~\ref{fig_rand_10_budget_scatter}-\ref{fig_block_10_budget_scatter}, we observe that the extent of correlation varies with different centrality metrics. To decide which centrality metric performs the best in describing the correlation, we adopt Pearson correlation~\add{\cite{Benesty2009}} to help us select the proper centrality metric. A perfect Pearson correlation $1$ occurs when each of the variables is a perfect monotone function of the other. On the contrary, $0$ means that there is completely no correlation between the two set of numbers. Table~\ref{centrality_corr_spearman} shows the result of Pearson correlation for Figs.~\ref{fig_rand_10_budget_scatter}-\ref{fig_block_10_budget_scatter}. From Table~\ref{centrality_corr_spearman}, we can see that PageRank performs the best except for the block-diagonal case because the definition of PageRank is infeasible for measuring the block-diagonal network. Fortunately, both the Eigenvector centrality and the Closeness centrality perform a strong linear correlation, which can be used for the block-diagonal case. 

From Figs.~\ref{fig_rand_10_budget_scatter}-\ref{fig_block_10_budget_scatter}, we find that there exists \add{a relatively strong correlation} between the investment/remaining work and its PageRank centrality under the four DSM architectures. So, as a quick heuristic, we can assign the budget as a function of centrality instead of solving a complex optimization problem, as previously found in \adddd{\cite{Braha2007,Maier2014}}\label{citemaier}. Moreover, PD managers can use PageRank centrality as a proxy to allocate development resources for the modules. 

\begin{figure*}
\centering 
\includegraphics[width=.975\linewidth]{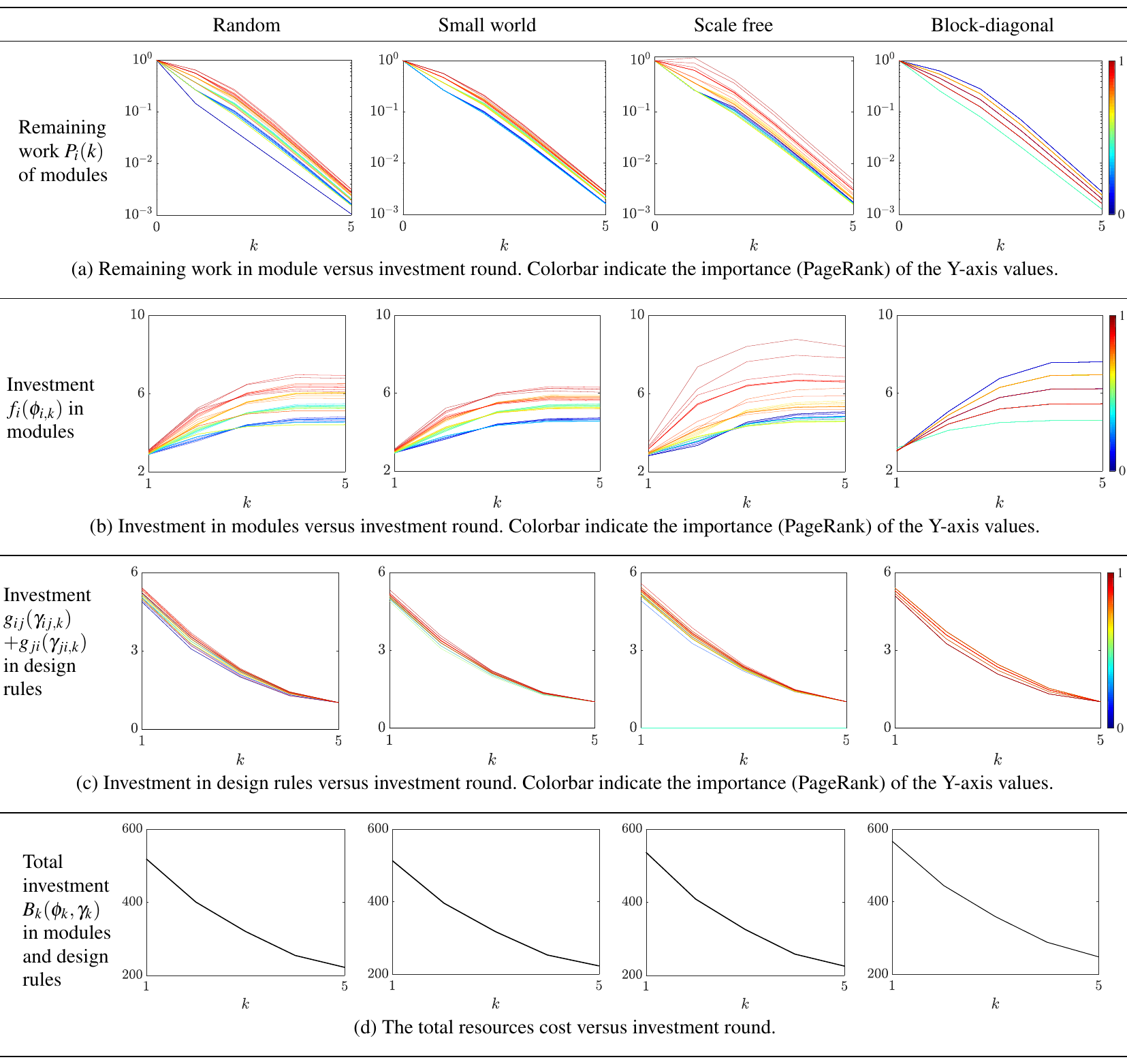}
\caption{The optimal solution of the performance-constrained problem. Color lines distinguish the importance of module/design rules via the PageRank.}
\label{result_work_constraint} 
\end{figure*}


For the correlation between the investment in the design rule and its centrality in Figs.~\ref{fig_rand_10_budget_scatter} (c)-\ref{fig_block_10_budget_scatter} (c), we can see that the extent of correlation in Figs.~\ref{fig_rand_10_budget_scatter} (c)-\ref{fig_scalefree_10_budget_scatter} (c) is not strong enough for drawing a conclusion. However, for the block-diagonal case in Fig.~\ref{fig_block_10_budget_scatter} (c), we observe a different phenomenon that hundred of variables are overlapped to a few points, and are strongly correlated with the Eigenvector and the Closeness centrality. Specifically, we notice that design rules belonging to different sub-blocks but with the same size receive the same amount of investment. In other words, the investment in design rules is independent of the block to which it belongs. This independence is caused by the special structure of the block-diagonal network, in which all the sub-blocks are independent of each other. 


\subsection{Analysis and discussion of the performance-constrained problem}\label{sim_prob2}

In this subsection, we solve the performance-constrained problem via convex optimization. Although we have revealed the trends of the decision variables and the internal relations for the budget-constrained problem, we cannot conclude that the same situation also exists in the performance-constrained problem.

As in Section~\ref{sim_prob1}, we perform the simulation experiment on a controlled set of product architectures.
For initializing the performance-constrained problem, we adopt the same parameters setting as the budget-constrained problem. Based on the formulation of the performance-constrained problem,
we set \add{the} constraint for the total remaining work of the final investment round to $\bar{P}=0.01$, which can be regarded as a threshold for judging the accomplishment of the PD process. For example, suppose that the total remaining work at the beginning is normalized to $1$, if we set $\bar{P}=0.01$, it means that when the total remaining work is $1\%$ its initial value, we can say that the project is finished. We \add{conduct} the experiments with the selected DSM architectures in Fig~\ref{figDSM}, and \add{observe} the following response variables: the remaining work in modules, the investment in the modules and the design rules, \add{the dependency strength between modules, and the total investment at each round.} 

Fig.~\ref{result_work_constraint} shows the evolution of decision variables and the remaining work of the performance-constrained problem. From Figs.~\ref{result_work_constraint} (a)-(c), we can see that the solution of the performance-constrained problem exhibits similar trends to the budget-constrained problem. Particularly, the solution shows that the product architecture evolves from an integral to a modular as successive investment are made on the modules and the design rules.
In Fig.~\ref{result_work_constraint} (d), it is worth noting that there is a decreasing tendency on the total investment during the PD process, which contradicts our intuition that the resources should be equally allocated for each investment round. In the performance-constrained problem, we also find the positive correlation between the investment in the module and its related design rules as the budget-constrained problem. \add{Further analysis similar to the ones for the performance-constrained problem} allows us to draw the same set of conclusions as the one we obtained for the budget-constrained problem. The details are omitted.

\begin{figure*}[tb]
\centering
\includegraphics[width=.975\linewidth]{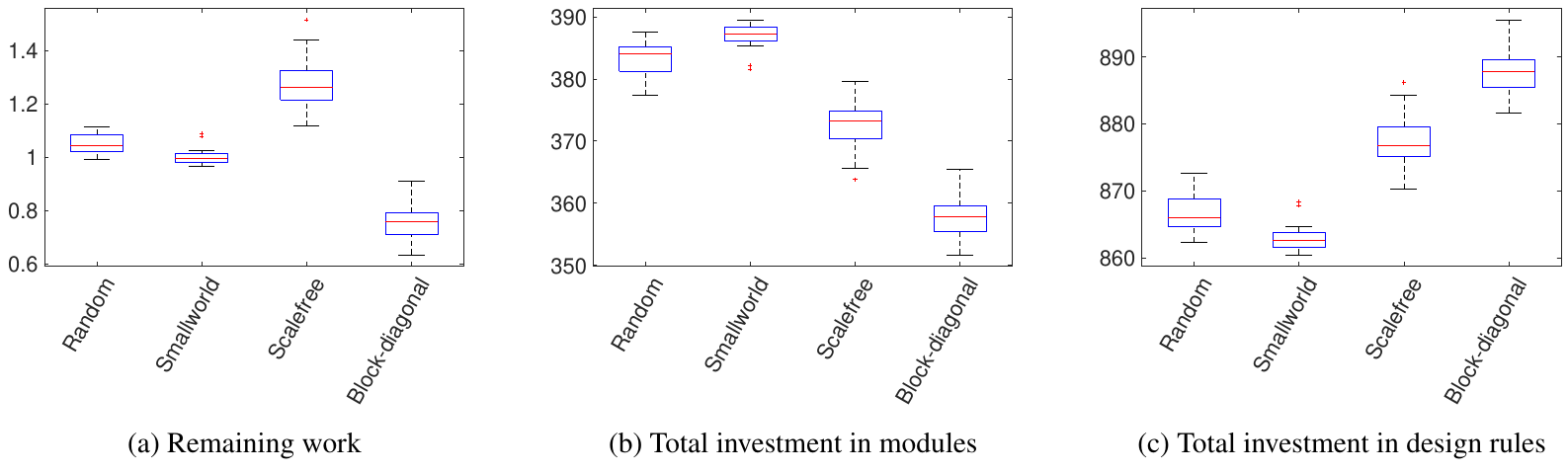}
\caption{The total investment and performance of the budget-constrained problem versus different DSM architectures. \adddd{The difference among the networks is statistically significant at the significance level of~10\%, except between the random and small world networks.}} 
\label{fig_budget_DSM_compare} 
\end{figure*}

\begin{figure*}[tb]
\centering
\includegraphics[width=.975\linewidth]{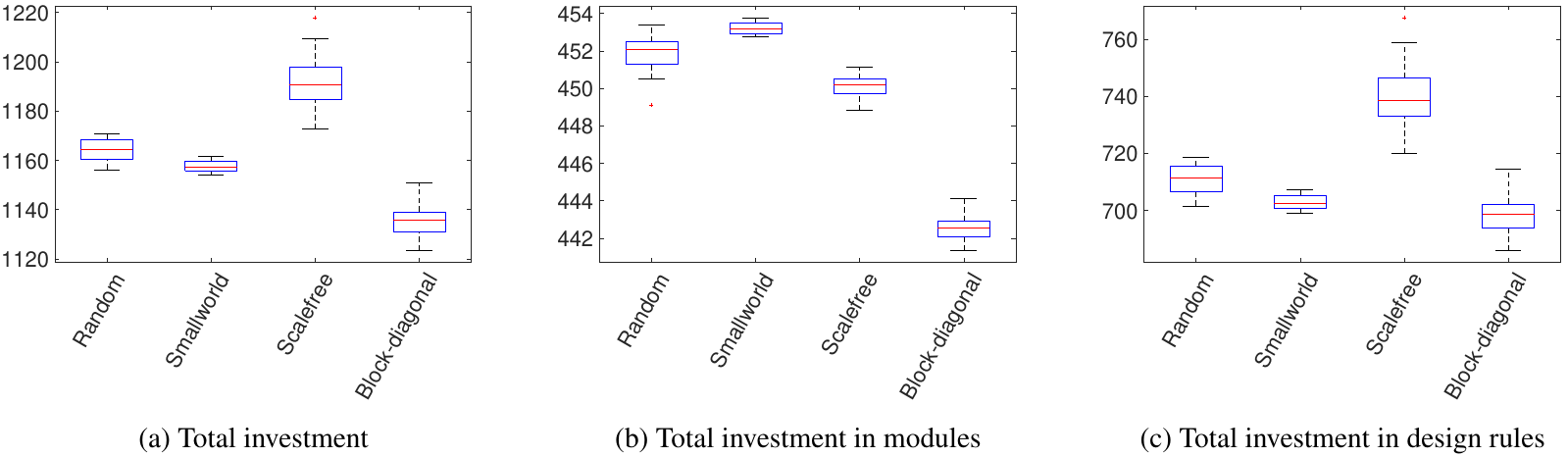}
\caption{The total investment and performance of performance-constrained problem versus different DSM architectures. \adddd{The difference among the networks is statistically significant at the significance level of~10\%, except between the random and small world networks.} } 
\label{fig_workconstraint_DSM_compare} 
\end{figure*}

\subsection{Analysis of different DSM architectures}
\label{sim_result_compare}

In this subsection, we carry out an analysis of variance on the product architecture to investigate whether the product architecture affects the resource allocation and the performance of the designed PD system. \add{It is important to remark that the set of synthetic networks we use for our analysis is not intended to replicate all the aspects of real design networks, specifically the significant difference between the distribution of in- and out-degrees in the product architecture~\cite{Braha2004,Braha2004a,Braha2007}.\label{context}}

In this experiment, we used the four DSM architectures introduced in Fig.~\ref{figDSM} and selected the three response variables: total remaining work, total investment in modules, and total investment in design rules. To detect any statistical difference, we randomly generate $50$ sample networks for each type of product architecture. In all the problems, we unify the parameters of the WTM and the cost functions as in the previous sections. We solve these problems with the proposed framework in this paper. The results of the budget-constrained problem and the performance-constrained problem sorted by product architecture are shown in Figs.~\ref{fig_budget_DSM_compare} and~\ref{fig_workconstraint_DSM_compare}, respectively. We use Boxplots to illustrate the maximum, the minimum, the variance, and the mean value of the investment and the remaining work. We observe that the architecture affects the resource allocation, which in turn affects the remaining work, investment in modules and design rules of the PD process. To statistically investigate the dependence, we also carry out one-way ANOVA tests~\cite{Tabachnick2007} between pairs of data. We adopt the $p$-value from the ANOVA test as an index to illustrate the difference between each two networks. \adddd{We observe that the difference among the networks is statistically significant at the significance level of~10\%, except between the random and small world networks.} This tendency could be partly attributed to the similarity in the construction rules for these two networks.\label{adddd}

For the remaining work of the budget-constrained problem (Fig.~\ref{fig_budget_DSM_compare} (a)), we can see that the block-diagonal architecture has the minimum remaining work (the best performance) compared with the other three architectures, which indicates that the modular architecture performs better than integral architecture (i.e., small world and scale free). Besides, we also notice that although the block-diagonal has the best performance, the variability is larger than the Small world case. This result implies that the stability of product architecture can not be neglected in designing the DSM structure. From Figs.~\ref{fig_budget_DSM_compare} \add{(c)} and \add{(e)}, we confirm that there exists a variance on the investment in modules and design rules with different product architectures. Also, from Fig.~\ref{fig_workconstraint_DSM_compare} for the performance-constrained problem, we can see that for meeting the same target of the remaining work, the block-diagonal architecture costs the minimum resources among the four architectures, which also indicates that the modular architecture performs better than integral architecture.

\begin{table*}[tb]
\centering
\begin{minipage}[t]{.49\linewidth}
\centering
\caption{Task duration of tire and wheel manipulator design}
\label{tab:case_task_1}
\includegraphics[width=.95\linewidth]{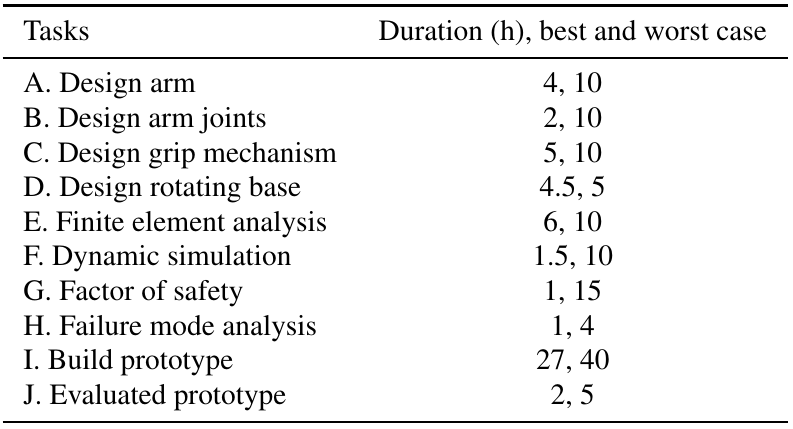}
\end{minipage}
\begin{minipage}[t]{.49\linewidth}
\centering
\caption{The WTM of the manipulator design process}
\label{tab:WTM_case_1}
\includegraphics[width=.9\linewidth]{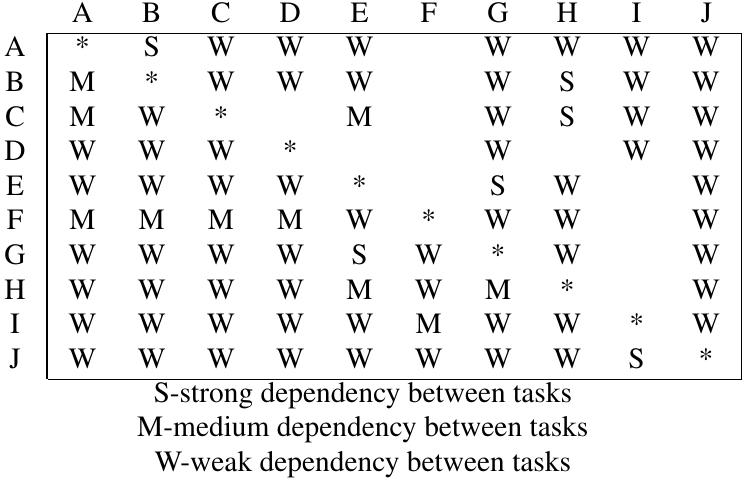}
\end{minipage}
\vspace{5mm}
\caption{WTM for Automotive appearance design}\label{tab:WTM_case_2}
\includegraphics[width=.575\linewidth]{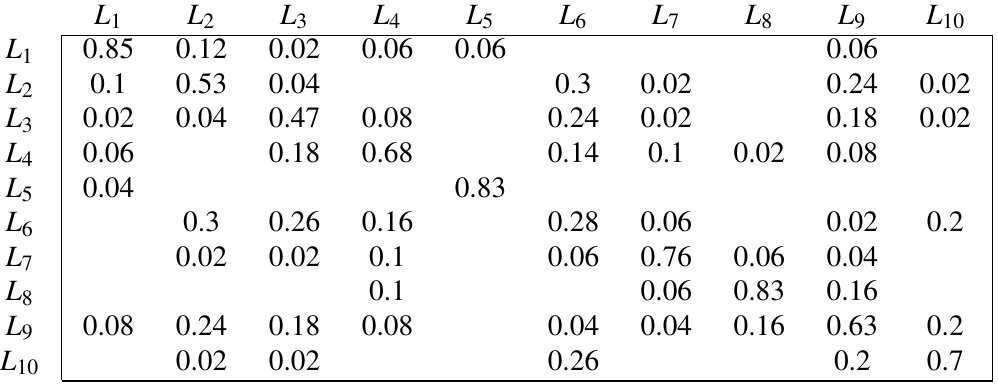}
\end{table*}

\subsection{Limitation and robustness}
\label{robustness} 

\add{We discuss the limitation and robustness of our analysis in this section. First, as we remarked in the introduction, the set of synthetic networks we use for our analysis does not necessarily cover all the types of realistic design networks.} \add{As revealed in the seminal works in~\cite{Braha2004,Braha2004a,Braha2007}, empirical product development architectures often exhibit distinctive asymmetry between the distributions of incoming and outgoing links. For this reason, we investigate real product development processes having asymmetric architectures in the next section.}\label{context_section} 

\add{Second, our findings above are dependent on the the current choice of relevant parameters. In order to examine the robustness of our analysis to the choice of parameters, we let the shape parameter $p$ of the cost function~\eqref{cost:example} vary between $[1, 10]$ and perform simulations. We then found that the solution (i.e., the resulting pattern of resource allocation) exhibits the same trend, which suggests the robustness of our analysis. We also vary the other shape parameter~$c_{ij}$ to find that, the larger $c_{ij}$, the lower the effect of the investment on the development process due to the higher price of the resources.}

\section{Case study}\label{case_study}

In this section, we demonstrate and validate our proposed model using two case studies considering the resource allocation problems for the PD processes presented in~\cite{George2012,Yassine2003}. 

\begin{figure*}[tb]
\centering
\includegraphics[width=.975\linewidth]{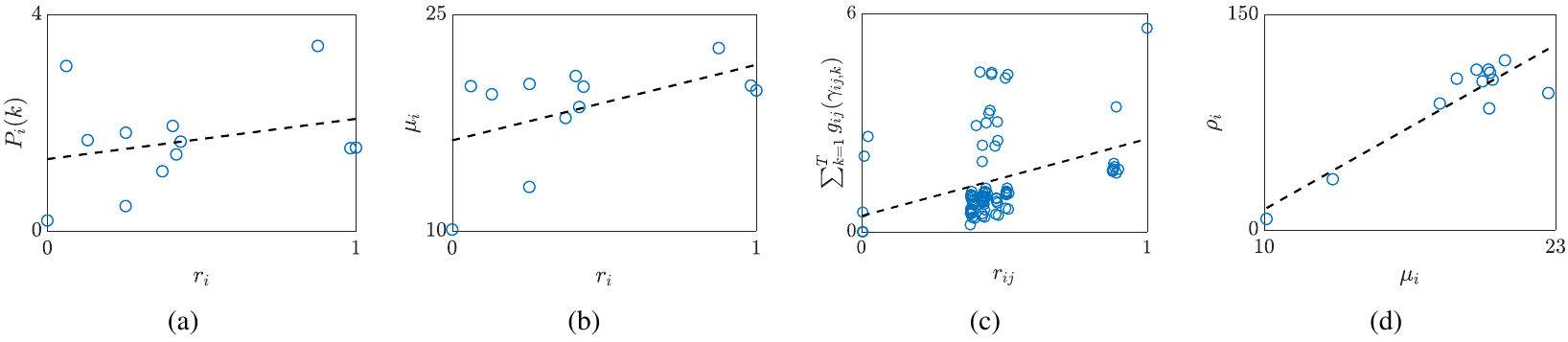}
\caption{Optimal investment in modules and design rules of the budget-constrained problem versus PageRank centrality in the tire and wheel manipulator design. (a) Remaining work versus centrality; (b) investment in modules versus centrality; (c) investment in design rules versus centrality; (d) investment in module versus the total investment in its dependent design rules.} 
\label{casestudy_1_work} 
\end{figure*}

\begin{figure}[tb]
\centering
\includegraphics[width=.975\linewidth]{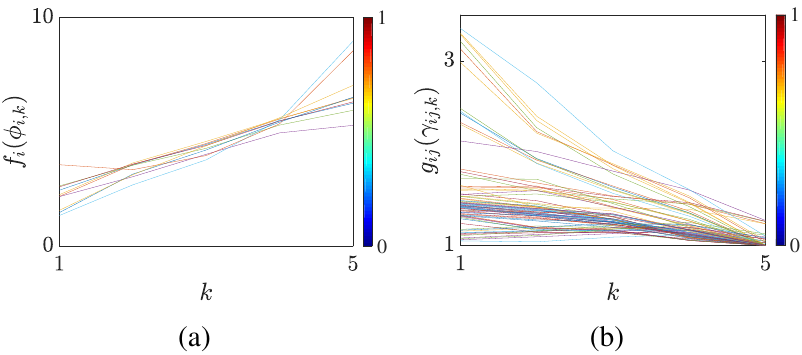}
\caption{The optimal solution of the budget-constrained problem of case study $1$. Color lines distinguish the importance of module/design rules via the PageRank. (a) Investment in modules versus centrality. (b) Investment in design rules versus centrality. } 
\label{casestudy_1_work_tradeoff} 
\end{figure}

\subsection{Case 1: Mechanical manipulator design}
\label{case_1}

In this section, we consider the case study problem \add{of} the tire and wheel manipulator design process presented in~\cite{George2012}. Table~\ref{tab:case_task_1} lists all the tasks for the design process and the best/worst duration \add{for each task} (in hours) per investment round.
The dependency structure of the design process, illustrating the interactions among the tasks, is demonstrated in the DSM (or WTM) \add{as} shown in Table~\ref{tab:WTM_case_1}. In Table~\ref{tab:WTM_case_1}, the diagonal entries represent the completion rate for handling the remaining work of the task, while the off-diagonal element \add{indicates} the ratio of the fractional work exchanged among the tasks. In this paper, we adopt the inverse of average duration as the value of the diagonal entries of WTM. Table~\ref{tab:WTM_case_1} shows the value of dependency strength between the tasks which are assigned in three numerical values: 0.5, 0,25 and 0.05 for strong (S), medium (M), and weak (W) dependencies, respectively. 

\add{The assessment of the various input parameters to our model can be easily collected through a set of structured interviews (perhaps using a survey instrument) following a similar procedure to the one described in~\cite{Yassine2016}. For the assessment of the various DSM parameters used in our analysis, we recommend a three-step procedure: Kick-off, individual estimation, and review/wrap-up. In the first step, a one-hour kick-off meeting is conducted with the product development project team to identify all the development tasks. In the second step, short individual interviews are conducted with each team member to solicit their estimate for task durations using two- or three-point estimates for most-likely duration, worst case duration, and most optimistic (best) duration. Also, in this step, interviewees are asked about their dependencies on other tasks from the already established list of tasks. A simple subjective scale can be used for the strength of the identified dependencies: strong, medium and weak. In the final step, the product development project team is assembled again to review the results of the interviews and resolve any discrepancy or disagreement in any of the estimates.}

In our case study, we specify the manipulator design problem with an initial work vector $P(0)=\add{[1,\, \dotsc,\, 1]^\top}$ (i.e., all the tasks at the beginning of the design process have $100$$\%$ work remaining). We adopt the mean value of the best and worst duration as a measure of the complexity of each task \add{(i.e., $A_1(\bar{\phi}_1, \bar{\gamma}_1)_i=(t_{\max{}}^i+t_{\min{}}^i)/2$)} the initial value of the diagonal entries in WTM). While the values of the dependency strength between tasks are represented by the off-diagonal entries of WTM. For the cost function, we unify the parameters of all cost functions with $c_{ij}=1$, $p=1$, $\Omega_{ij}=1$, in that case, the unit of investment in the individuals has the same effect. Also, we let the variable $\phi$ and $\gamma$ of cost function vary within the interval $[0.1, 1]$, which means that, after each investment round, the rate of dealing process in each module and design rule can be accelerated between $[0$$\%-90$$\%]$. Also, we set investment round $T=5$ for the PD process. In the budget-constrained problem, we let the investment budget for each round equally fixed by $\bar{B}_k=200~(k =1, \dotsc, T)$. For the performance-constrained problem, we set the target for the remaining work $\sum_{i=1}^n P_i(T)=0.01$, which means that if the remaining work exceeds the threshold $1$$\%$, we can assume the completion of the PD process. From Section~\ref{simulation}, we have shown the trends and structure of the optimal solution, so in the case study, we only focus on analyzing the relations between the importance of modules and design rules (i.e., their centrality) in the WTM network and their allocated investment/performance. 

\begin{figure*}[tb]
\centering
\includegraphics[width=.975\linewidth]{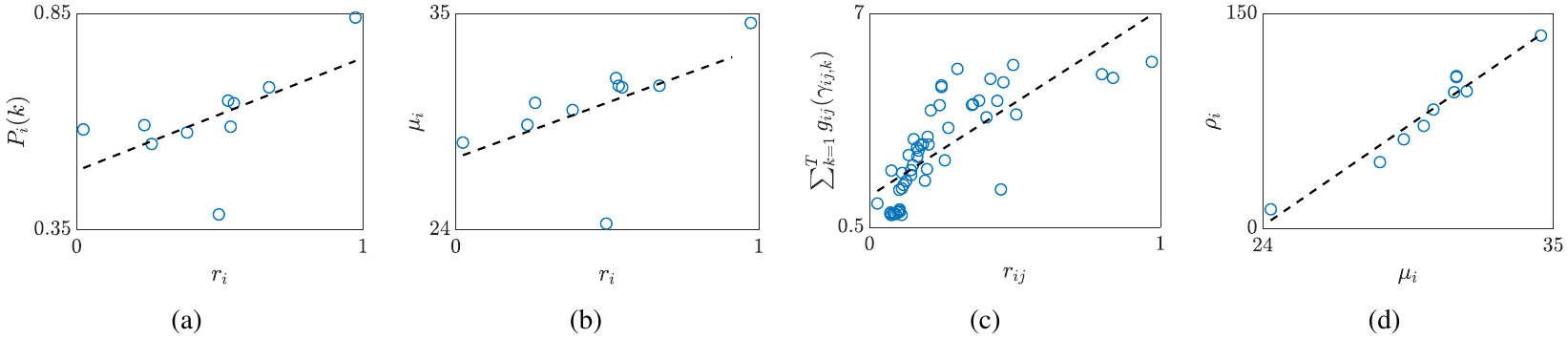}
\caption{Optimal investment in modules and design rules of the budget-constrained problem 
versus PageRank centrality in the automotive appearance design. (a) Remaining work versus centrality; (b) investment in module versus centrality; (c) investment in design rules versus centrality; (d) investment in module versus the total investment in its dependent design rules.}

\label{casestudy_2_work} 
\end{figure*}

\begin{figure}[tb]
\centering
\includegraphics[width=.975\linewidth]{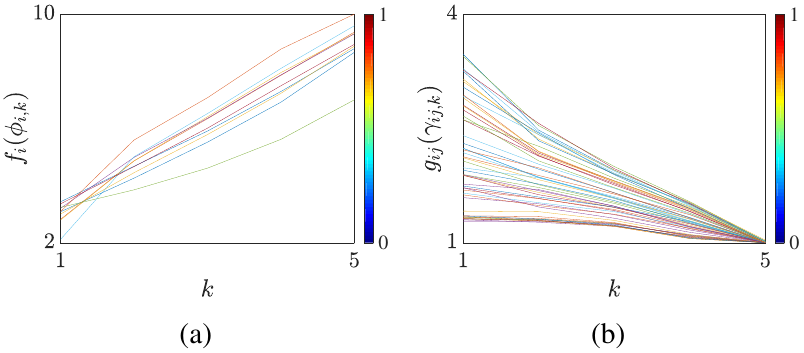}
\caption{The optimal solution of case study $2$. Color lines distinguish the importance of module/design rules via the PageRank. (a) Investment in modules versus centrality. (b) Investment in design rules versus centrality.
}
\label{casestudy_2_work_tradeoff} 
\end{figure}

Based on the analysis result from Section~\ref{simulation}, we adopt the PageRank centrality for measuring the importance of the task/dependency in the WTM network. From Fig.~\ref{casestudy_1_work}, we can see that both problems reveal the same trends and structure. In Fig.~\ref{casestudy_1_work} (a), the regression line shows that there is a correlation but not strong between the remaining work in the module and its PageRank centrality. We can still conclude that the task with higher centrality tends to have a larger amount of remaining work. In Fig.~\ref{casestudy_1_work} (b), we get nearly the same trend that the task with higher centrality tends to receive larger investment. However, in Fig.~\ref{casestudy_1_work} (c), there is still an ambiguous correlation between the investment in dependency and its corresponding centrality, which coincides with the analysis result in Section~\ref{simulation}. From Fig.~\ref{casestudy_1_work} (d), we can see there is a clear positive correlation between the investment in the task and the total investment in its dependencies. \add{In Fig.~\ref{casestudy_1_work_tradeoff}, we present the evolution of investment in modules and design rules. We can confirm the increasing trend in modules and decreasing trend in design rules, as we have found in our synthetic analysis in Section~\ref{simulation}.}

\subsection{Case 2: Automotive appearance design}
\label{case_2}

In this section, we adopt the automotive appearance design presented in~\cite{Yassine2003}, which focuses on the design process of the interior and the exterior surface of the automotive for achieving a balance among the appearance, quality, and operational interface. However, we only focus on the local team which contains the following tasks: 1) carpet, 2) center console, 3) door trim panel, 4) garnish trim, 5) overhead system, 6) instrument panel, 7) luggage trim, 8) package tray, 9) seats, and 10) steering wheel.

The WTM for the appearance design is given in Table~\ref{tab:WTM_case_2} where the diagonal entries represent the completion rate on each module, while the off-diagonal entries represent the ratio of the rework exchanged between each task. In this case study, we specify the automotive appearance design problem with an initial work vector $P(0)=\add{[1,\, \dotsc,\, 1]^\top}$. For the cost function, we unify the parameters of all cost functions with $c_{ij}=1$, $p=1$, $\Omega_{ij}=1$, in that case, the unit of investment in the individuals has the same effect. Also, we let the variable $\phi$ and $\gamma$ of cost function vary within the interval $[0.1, 1]$, which means that, after each investment round, the the completion rate on each module and design rule can be accelerated between $[0$$\%-90$$\%]$.

We use 5 investment rounds $(T = 5)$ for the design process. In the budget-constraint problem, we let the amount of investment at each round equally fixed to $\bar{B}_k=200$. For \add{the performance-constraint problem}, we set the target for the remaining work $0.01$, which means that if percentage of the remaining work is $1$$\%$ of its initial value, then this will indicate the completion of the PD process. As in the previous case study, we focus on analyzing the relations between the importance of modules and design rules in the WTM and their allocated investment/performance. 

Fig.~\ref{casestudy_2_work} shows the investment versus the PageRank centrality measures. For the relationship between the performance/remaining work of the module and its importance, we can see from Fig.~\ref{casestudy_2_work} (a) that the module with a higher centrality has a larger amount of remaining work. This result coincides with our intuition that the module has a lower centrality means it has higher efficiency, which naturally results in the lower level of remaining work. Similarly, in Fig.~\ref{casestudy_2_work} (b), we can conclude that the module with higher centrality (i.e., lower efficiency) gets the larger investment for improving its performance. As for the design rules, Fig.~\ref{casestudy_2_work} (c) shows the relation between the investment in design rules and its centrality in the WTM network. From the linear regression line, we can see that the correlation is not very strong, but there is still a tendency that the design rule with the higher edge centrality (i.e., connects the two lower efficiency modules) tend to attract more resources. Finally, Fig.~\ref{casestudy_2_work} (d), we can see that there is a strong correlation that the investment in the module with higher centrality (i.e., lower efficiency) coincide with the total investment in its dependencies. \add{Finally, in Fig.~\ref{casestudy_2_work_tradeoff}, we present the evolution of investment in modules and design rules.
As we found in the first case study, we can confirm the trade-off between the investment in modules and design rules as increasing trend in modules and decreasing trend in design rules.}


\section{Conclusion and discussion}
\label{conclusion}

PD managers are always faced with the problem of making resource allocation decisions. Especially, when there are several investment rounds, making optimal decisions based on intuition or heuristic rules becomes rather difficult. Although, some literatures have proposed analysis methods
on complex network theory~\add{\cite{Braha2004,Braha2004a,Batallas2006,Braha2007,Collins2009}}\label{chron} and dynamic linear systems~\cite{Smith1997,Yassine2003}, they lack of an analytical and mathematical
optimization framework similar to the one presented in this
paper. 

Our results provide PD managers with an efficient tool to allocate development resources optimally for \add{the budget-constrained problem and performance-constrained problem}, where the resources can be allocated on both modules and design rules.
Although we carried out the experiments with two types of problems, and with different product architectures for each problem, the evolution of the investment and remaining work exhibit similar trends, which shows that the evolution property of the PD process is independent of the problem formulation and product architecture. Moreover, the investment and performance in modules also illustrate that certain correlations exist despite the problem formulation and product architecture, which also confirms that these trends and correlations are the intrinsic properties of the PD process.

In the analysis of different PD architectures, we show that the architecture of the product affects resource allocation which in turn affects the performance of the PD process. Design and managerial guidelines can result from the direct analysis of the PD architecture. Specifically, for development engineers, our result can be used for selecting the product architecture which leads to maximum performance. On the other hand, when the PD architecture is fixed, our proposed framework helps PD managers in deciding on the optimal budget proportions to be allocated to modules and to design rules.

Furthermore, for making a further utilization of our framework, we discuss the feasibility of adding the linear regression lines for exploring the possibility of making resource allocation decisions without the need for solving the optimization problem (i.e., directly by utilizing the correlation results). In other words, our proposed framework allows us to gain insights into the relationship between the investment and the DSM architecture, which inspires us to further investigate and model the mapping from the centrality of a module to its investment. From the results in Section~\ref{sim_result_compare}, we know that the DSM architecture affects the resource allocation which in turn affects the performance. Specifically, from the regression lines in Figs.~\ref{fig_rand_10_budget_scatter}-\ref{fig_block_10_budget_scatter}, we observed that the slope in each figure varies among different DSM architectures, which indicates that modeling a general investment function for all the DSMs is not feasible. However, if we fix the DSM architecture and the problem size (i.e., number of modules and design rules), is it possible for us to address this problem? From the problem formulation in Section~\ref{Pb_formulation}, we know that there are numerous parameters in the WTMs and the cost functions that can affect the shape of the regression line. Currently, we cannot determine what DSM parameters influence the slope of the regression line. 


One limitation of the framework proposed in this paper is that it does not consider the time-delay effect; so, dynamic investment problem with time-delay need to be considered in future work; especially, if the parameters of the PD system are updated after a certain period. Then, the investment decision making problem becomes \add{applicable to} a more general situation.



%
%



\section*{Appendix}

\add{In this appendix, we illustrate how we can reduce Problems~\ref{pb1} and~\ref{pb2} to convex optimization problems. We introduce the following notations for cost functions. 
For the total cost function in \eqref{cost}, we define 
\begin{align*}
B_k^+(\phi_{k}, \gamma_{k})=\sum_{i=1}^{n} f_i^+(\phi_{i, k})+\sum_{i=1}^{n}\sum_{i\neq j}g_{ij}^+(\gamma_{ij, k}),
\\
B_k^-(\phi_{k}, \gamma_{k})=\sum_{i=1}^{n} f_i^+(\bar{\phi}_{i, k})+\sum_{i=1}^{n}\sum_{i\neq j}g_{ij}^+(\bar{\gamma}_{ij, k}).
\end{align*}}

\add{Let us first show that Problem~\ref{pb1} reduces to solving a convex optimization problem. Notice that the optimization problem~\eqref{pb1:} is equivalent to 
\begin{subequations}\label{pb1:minimize}
\begin{align}
{\minimize_{\phi, \gamma}} & \sum_{i=1}^{n}P_i(T) \label{pb1:minimize:A}
\\
\subjectto & B_k(\phi_{k}, \gamma_{k}) \leq \bar{B}_k, \label{pb1:minimize:B}
\\
&\add{0<\ubar{\phi}_{i, k} \leq \phi_{i, k} \leq \bar{\phi}_{i, k},}\label{pb1:minimize:D}\\ &\add{0<\ubar{\gamma}_{ij, k} \leq \gamma_{ij, k} \leq \bar{\gamma}_{ij, k},}~\add{k=1, \dotsc ,T.}\label{pb1:minimize:C} 
\end{align}
\end{subequations}
Under this notation, we can show that 
the solution of the budget-constrained problem is given by
\begin{equation}\label{thm1var}
\phi=\exp [x],~\gamma=\exp [y],
\end{equation}
where $\exp [\cdot]$ is the entrywise exponential function of the variables, and 
$x=\{x_k\}^T_{k=1}$ and $y=\{y_k\}^T_{k=1}$ solve the following \emph{convex} optimization problem: 
\begin{subequations}\label{thm1}
\begin{align}
\addd{\minimize_{x, y, \Gamma }} & \Gamma
\label{thm1:A}	\\
\subjectto & \log B_k^+(x_{k}, y_{k}) \leq \log (\bar{B}_k+B_k^-), \label{thm1:B}
\\
&	\log \add{\sum_{i=1}^{n}P_i(T)} \add{\leq \Gamma}, \label{thm1:C}\\
&\log \ubar{\phi}_{i, k} \leq x_{i, k} \leq \log \bar{\phi}_{i, k},\label{thm1:D}\\
&\log \ubar{\gamma}_{ij, k} \leq y_{ij, k} \leq \log \bar{\gamma}_{ij, k}.\label{thm1:E}
\end{align}
\end{subequations}}

\add{Let us give a brief proof of this statement. Under Lemma~\ref{lem1}, it can easily be seen that \eqref{pb1:minimize:A}, \eqref{pb1:minimize:B}, and~\eqref{pb1:minimize:C} in the budget-constrained problem are equivalent to \eqref{thm1:B}, \eqref{thm1:C}, \eqref{thm1:D}, and \eqref{thm1:E}. Therefore, the solution of the optimization problem~\eqref{thm1} given by \eqref{thm1var} is the solution of the budget-constrained problem. Under this equivalence, we show the convexity of the optimization problem~\eqref{thm1}. It is sufficient to show that constraints~\eqref{thm1:B} and~\eqref{thm1:C} are convex if the performance functions~\eqref{pb1:minimize:A}, \eqref{pb1:minimize:B} and the cost function~\eqref{cost} follow Definition~\ref{def1}.}

\add{We can similarly reduce Problem~\ref{pb2} to a convex optimization problem. We can specifically show that the solution of the performance-constrained problem is given by \eqref{thm1var}, where $x=\{x_k\}^T_{k=1}$ and $y=\{y_k\}^T_{k=1}$ solve the following convex optimization problem
\begin{subequations}
\begin{align*}
\minimize_{x, y, \Psi} & \Psi
\\
\subjectto & \log \add{\sum_{i=1}^{n}P_i(T)} \add{\leq \log \bar{P}}, 
\\
&	\log \sum_{k=1}^{T} B_k^+(x_{k}, y_{k}) \leq \log \left(\Psi+ \sum_{k=1}^{T} B_k^-\right) ,\label{thm2:C}\\
&\log \ubar{\phi}_{i, k} \leq x_{i, k} \leq \log \bar{\phi}_{i, k},
\\
&\log \ubar{\gamma}_{ij, k} \leq y_{ij, k} \leq \log \bar{\gamma}_{ij, k}.
\end{align*}
\end{subequations}
We omit the proof of this statement because it is similar to the one for the budget-constrained problem.
}

\end{document}